\documentclass[twocolumn]{aastex7}

\begin{document}

\title{Evidence of Gaia Enceladus experiencing at least two passages around the Milky Way\footnote{Based on observations collected at the European Southern Observatory under ESO programme IDs 109.22VP and 110.240W, as well obtained from the ESO Science Archive Facility.}}

\author[orcid=0000-0001-9155-9018]{Ása Skúladóttir}
\affiliation{Dipartimento di Fisica e Astronomia, Universit\'{a} degli Studi di Firenze, Via G. Sansone 1, I-50019 Sesto Fiorentino, Italy.}
\email[show]{asa.skuladottir@unifi.it}  

\author[orcid=0000-0001-6541-1933]{Heitor Ernandes} 
\affiliation{Lund Observatory, Department of Geology, Lund University, S\"olvegatan 12, Lund, 223\,62, Sweden.}
\email{heitor.ernandes@geol.lu.se}

\author[orcid=0000-0002-3101-5921]{Diane K. Feuillet}
\affiliation{Lund Observatory, Department of Geology, Lund University, S\"olvegatan 12, Lund, 223\,62, Sweden.}
\affiliation{Observational Astrophysics, Department of Physics and Astronomy, Uppsala University, Box 516, SE-751 20 Uppsala, Sweden.}
\email{fakeemail3@google.com}

\author[0009-0003-0816-2880]{Alice Mori}
\affiliation{Dipartimento di Fisica e Astronomia, Universit\'{a} degli Studi di Firenze, Via G. Sansone 1, I-50019 Sesto Fiorentino, Italy.}
\affiliation{INAF/Osservatorio Astrofisico di Arcetri, Largo E. Fermi 5, I-50125 Firenze, Italy.}
\email{fakeemail5@google.com}

\author[orcid=0000-0002-7539-1638]{Sofia Feltzing}
\affiliation{Lund Observatory, Department of Geology, Lund University, S\"olvegatan 12, Lund, 223\,62, Sweden.}
\email{fakeemail6@google.com}

\author[0000-0002-9750-1922]{Romain E.~R. Lucchesi}
\affiliation{Dipartimento di Fisica e Astronomia, Universit\'{a} degli Studi di Firenze, Via G. Sansone 1, I-50019 Sesto Fiorentino, Italy.}
\email{fakeemail7@google.com}

\author[0000-0002-5213-4807]{Paola Di Matteo}
\affiliation{LIRA, Observatoire de Paris, PSL Research University, CNRS, Place Jules Janssen, 92195 Meudon, France.}
\email{fakeemail7@google.com}

\begin{abstract}
One of the major recent breakthroughs has been the discovery of the last Major Merger to happen in the history of the Milky Way. Around 10~Gyr ago the galaxy Gaia Enceladus, with estimated $\sim$10\% of the Milky Way mass, fell into its potential, bringing a large amount of stars which can be identified through their unique chemical and kinematic signatures. Simulations have long predicted that a galaxy of this size should experience several passages through the disk of the Milky Way before eventually being fully dispersed. For the first time, we present observational evidence to support this. 
We identify two subpopulations accreted from Gaia Enceladus: 1)~stars which today have large kinematic energy, which originate from the outskirts of Gaia Enceladus and were accreted during early passages; 2)~stars with low kinetic energy accreted at later passages, originating from the inner parts of Gaia Enceladus.  
Through the use of high-precision chemical abundances, crucially including new aluminum measurements, we show that in all observed abundance ratios ([Fe/H], [Al/Fe], [Mg/Fe] and [Mg/Ba]), stars with high energy show evidence of coming from a less chemically evolved outer region of Gaia Enceladus, compared to the stars with low energy. We therefore conclude that Gaia Enceladus experienced several passages before merging with the main body of our Galaxy.
This discovery has wide implications for our understanding of this event, and consolidates Gaia Enceladus as a benchmark for studying galaxy mergers and hierarchical galaxy formation in extraordinary details. 
\end{abstract}


\section{Introduction} \label{sec:intro}

The last Major Merger of the Milky Way was first identified over a decade ago through a detailed study of chemical abundances and kinematics of stars in the solar neighborhood ($d<350$\,pc; \citealt
{Nissen97,Nissen10,Nissen11}; hereafter the NS sample). Unexpectedly, the stars with halo kinematics separated into two populations, with clear differences in the [$\alpha$/Fe] abundance ratios (e.g. [Mg/Fe]). This pioneering study interpreted the high-$\alpha$ population to be stars formed in situ in the Milky Way, and the low-$\alpha$ population as stars accreted from a dwarf galaxy. Recently, this discovery was spectacularly and undeniably confirmed by the Gaia space mission and the accreted galaxy was given the name {\it Gaia Enceladus}\footnote{also known as the Gaia Sausage.} \citep{Belukorov18,Helmi18,Haywood18}. Among the $\sim$100 stellar streams which have now been identified thanks to the high-quality Gaia photometry and astrometry \citep[e.g.][]{Ibata21}, Gaia Enceladus stands out as the most massive accreted structure. With a stellar mass of $M_\star\sim10^{9-10}$\,M$_{\odot}$ \citep[e.g.][]{Feuillet20} it is about 100 times more massive than all other known Galactic mergers combined.

The interpretation of the low-$\alpha$ population in the NS sample being formed in a smaller galaxy was originally based on observations which consistently show dwarf galaxies to have lower [$\alpha$/Fe] ratios at a given [Fe/H], compared to the Milky Way \citep[e.g.][]{Tolstoy09}. Abundance ratios such as [Mg/Fe] and [Mg/Ba] are powerful tracers of star formation in different systems, since they are very sensitive to timescales \citep[e.g.][]{Ernandes24}. The element Mg is created by core-collapse supernovae (ccSN) on short timescales ($\sim10^7$\,Gyr), while Fe is also formed in Type Ia supernovae (SNIa, $\sim10^9$\,Gyr) and Ba is primarily formed in asymptotic giant branch (AGB) stars ($\gtrsim10^8$\,Gyr). Therefore, both [Mg/Fe] and [Mg/Ba] are typically lower in stars formed in an environment that experienced less efficient star formation and chemical enrichment \citep[e.g.][]{Tolstoy09,Skuladottir20a,Skuladottir20}. 

Similarly, the [Al/Fe] abundance ratios can be used to distinguish between stars formed in situ and those formed in a smaller galaxy, which later merged with the Milky Way \citep{Hawkins2015,Das2020,Horta2021,Feuillet21,Feltzing23}. In general, a smaller galaxy will have experienced a slower star formation which results in lower [Al/Fe]. Differences in elemental abundance ratios are thus found between different galaxies \citep[e.g.][]{Tolstoy09}, but also between different regions of galaxies which have experienced a variation in their chemical enrichment histories (e.g.~\citealt{Hayden15}; Lucchesi et al. in prep.).

Last year, the NS halo sample was reanalyzed with improved stellar atmospheric parameters, revealing that the accreted low-$\alpha$ stars above $\rm[Fe/H]>-1$ split into two subgroups based on their [Mg/Fe] ratios and kinematics~\citep{Nissen24}. The authors suggested that the accreted stars might come from two separate merger events, Gaia Enceladus and Thamnos \citep{Koppelman19}. However, \citet{Matsuno24} found that the kinematics of the second group was not in good agreement with Thamnos and instead suggested that these stars belonged to~Eos. The Eos structure was first identified by \citet{Myeong22}, and was proposed to have formed in situ, as evident by the observed high [Al/Fe] abundance ratios. However, no Al abundances were available for the NS sample to verify this suggestion. Clearly, the origin of this separation in [Mg/Fe] is still debated. Therefore, we investigate here for the first time the hypothesis that this separation is not showing two separate mergers but instead providing evidence of two groups of stars that have been stripped from Gaia Enceladus at different passages through the Milky Way potential. With a combination of statistical, chemical and kinematical arguments we aim to identify the stars that were accreted in the early passages of the Gaia Enceladus, versus those that were accreted later. 

\begin{figure*}
\begin{center}
\includegraphics[width=0.95\hsize]{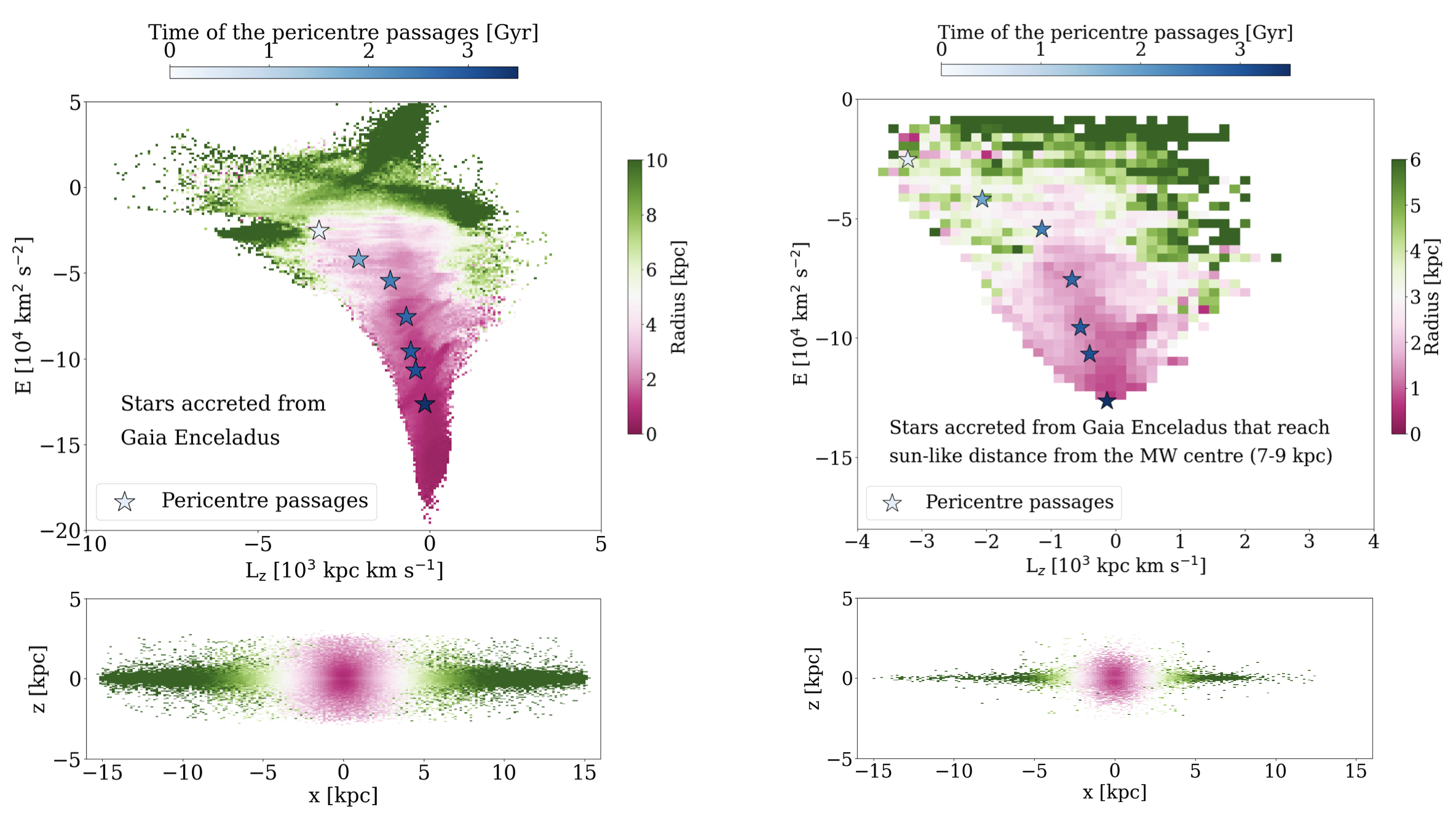} 
\caption{Simulation of Gaia Enceladus falling into the Milky Way \citep{Mori24}. 
Top panels show the energy and angular momentum ($E-L_z$) relation for the accreted stars (small points) which are now residing in the Milky Way. Left panel shows all stars, and the right panel those at a sun-like distance from the Galactic center (7-9\,kpc). Color-coding at a given point in ($E-L_z$) shows the average position of stars in Gaia Enceladus before the merger - the outskirts in green, and the inner regions in pink, as shown by the bottom panels. Star symbols (white to blue) show how the barycenter of Gaia Enceladus loses energy and angular momentum with time and more passages through the Milky Way.
}
\label{fig:am}
\end{center}
\end{figure*} 

\section{Data and Simulations} \label{sec:datamodel}

\subsection{Observational data}\label{sec:data}
In this work we present new measurements of aluminum (full analysis of the spectra will be presented in Ernandes et al. in prep.) in a subsample of the historically well-studied Galactic halo sample in the solar neighborhood (\citealt{Nissen10,Nissen11,Nissen24}; the NS sample). Other elemental abundances, i.e. the [Mg/Fe] and [Mg/Ba] in Sec.~\ref{sec:abu}, were adopted from \citet{Nissen11} and \citet{Nissen24}. We also adopt their classification of the halo sample into high- and low-$\alpha$ stars, and include as well their thick disk stars, i.e. with velocities $V_{tot}<\rm 180\,km\,s^{-1}$ \citep[][see their Fig.~3]{Nissen10}. All results of our data analysis are given in an online Table~\ref{tab:Al}. 

The new observations for this study come from two ESO VLT/UVES programs (PI:~Sk\'{u}lad\'{o}ttir, ESO Programme IDs: 109.22VP and 110.240W). The spectra are of high-resolution ($R\geq40\,000$) and have high signal-to-noise ratios, $\textsl{SNR}\geq150$\,pix$^{-1}$. Since the new observations do not include the entire NS sample, the ESO archive was searched for spectra of comparable quality for the missing stars. This led to spectra for 8 additional low-$\alpha$ stars to be added to the Al analysis. In total, we present Al measurements for 49 stars from the original NS sample.

All stellar parameters were adopted from a previous high-quality study of the sample \citep{Nissen24}, after verifying that their results were consistent with the analysis of our new spectra. To determine the elemental abundance we perform a standard spectral synthesis, using TURBOSPEC \citep{Alvarez98,Plez12}, with MARCS model atmospheres \citep{Gustafsson08} assuming 1D~and local-thermodynamical equilibrium (LTE). The {Al}\,{\sc i} line at 3961\,\AA\ was used for the elemental abundances determination. Although the Al\,{\sc i} line at 3944\,\AA\ was too blended to be trustworthy for a high-precision measurement, it was generally in good agreement with the abundance derived from the 3961\,\AA\ line. For our typical stellar parameters, we found the non-LTE effects on the 3961\,\AA\ Al line to be negligible ($\lesssim0.02$\,dex). These tests were done using the upgraded version of TURBOSPECTRUM \citep{Gerber23,Storm23} to do a full non-LTE fitting of the Al line as recommended by \cite{Nordlander17}, see more details in Ernandes et al., in prep. 

For aluminum we estimate a typical precision error of $\rm \Delta[Al/Fe]=0.10\,dex$. Since all spectra are of similar high quality and of stars with similar stellar parameters (see~\citealt{Nissen24}), the size of the error is very stable across the sample. This error estimation is supported by the scatter of the data, where the standard deviation of the high-$\alpha$ sample (including thick disk stars) is $\sigma_{h\alpha}=0.11$\,dex, and $\sigma_{l\alpha}=0.06$\,dex for the low-$\alpha$ sample. Note that these scatter measurements are effected both by the measurement errors as well as the intrinsic scatter and/or abundance trends. See full details on the error analysis in Ernandes et al. in prep. The error on [Mg/Fe] is significantly smaller, as seen by the small scatter in the high-$\alpha$ population, $\sigma_{h\alpha}=0.04$\,dex \citep{Nissen24}, which we adopt as a typical value.\footnote{Given the clear declining trend of [Mg/Fe] in the low-$\alpha$ population, this will greatly contribute to the scatter, $\sigma_{l\alpha}=0.08$\,dex, making it less reflective of measurement errors.} Through similar analysis we adopt a conservative $\rm\Delta[Mg/Ba]=0.10\,dex$, as it reflects the scatter in the low-$\alpha$ population \citep {Nissen11}. When calculating the averages or trends of abundances in Sec.~\ref{sec:abu} we therefore treat all measurements equally, not giving different weights to different stars.

Energies and angular momenta of the stars were calculated using \texttt{galpy} \citep{Bovy15} with distances (\texttt{distance\_gspphot}) and proper motions taken from {\it Gaia} Data Release~3 \citep{Gaia16, Gaia23} and radial velocities taken from \citet{Nissen10} and \citet{Nissen11}. We assume \texttt{MWPotential2014} \citep{Bovy15} as the Milky Way potential model and use the \texttt{actionAngleStaeckel} approximation \citep{Bovy2013, Binney2012} with a delta value of~0.4. The mean uncertainties of our input parameters are $0.7$\,pc in distance, $0.03$\,mas\,year$^{-1}$ in both RA and Dec. proper motion, and $0.3$~km~s$^{-1}$ in radial velocity. The resulting uncertainties in kinematics are expected to be approximately $0.12 \times 10^3$\,kpc\,km\,s$^{-1}$ in angular momentum and $0.28 \times 10^4$\,km$^2$\,s$^{-2}$ in energy based on the Monte Carlo analysis of \texttt{Gaia} parameter uncertainties done by \citet{Feuillet20}.

\subsection{Simulations} \label{sec:model}

The simulations we rely on in this paper have been used in previous studies \citep{Pagnini23,Mori24} and are similar to those presented in \citet{Jean-Baptiste17}. The analysis is based on a dissipationless, self-consistent, high-resolution N-body simulations of a Milky Way-type galaxy, during its accretion of a Gaia Enceladus-type satellite galaxy, with a mass ratio of 1:10. The main galaxy and its satellite are modeled as a collection of particles experiencing tidal
effects and dynamical friction, in a fully self-consistent manner. The accretion event is followed for 5\,Gyr, allowing for dynamic relaxation. Both galaxies are embedded in a dark matter halo, and contain a thin, an intermediate and a thick stellar disc – mimicking the Galactic thin disc, the young thick disc and the old thick disc, respectively \citep{Haywood13,DiMatteo16}. The barycenter of the satellite galaxy is followed both in position and velocity from an initial distance of 100\,kpc from the Milky Way-type galaxy. A more complete description of the simulations is available in \citet{Mori24}.

\begin{figure}
\begin{center}
\includegraphics[width=0.9\hsize]{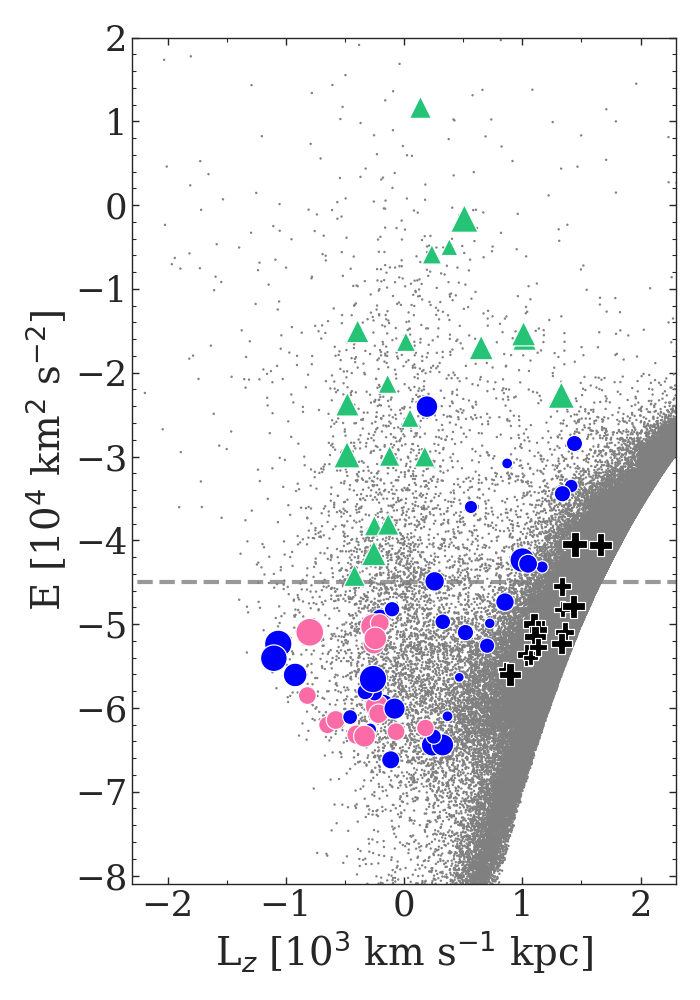} 
\caption{Energy and angular momentum ($E-L_z$) relation of observed stars. Accreted stars (low-$\alpha$) are separated into high and low energies, shown with green triangles and pink circles, respectively.
Thick disk stars are shown with black crosses, high-$\alpha$ stars with blue circles. The point size increases with decreasing [Fe/H]. For comparison, Milky Way field stars from APOGEE Data Release~17 are shown with small gray dots 
\citep{Majewski17,Abdurrouf22}, selected as described in \citet{Feltzing23}.
}
\label{fig:sel}
\end{center}
\end{figure}

\section{High and Low Energy stars} \label{sec:energy}

Simulations show clearly that stars from a major merger can cover a large range of dynamics \citep{Jean-Baptiste17,Amarante22,Khoperskov23b,Mori24}, with discrete over-densities in the energy-angular momentum ($E-L_z$) plane corresponding to stars stripped during different passages of the merging galaxy through the Milky Way potential, see Fig.~\ref{fig:am} (left). A possible signature of this has been reported in the Gaia data \citep{Belokurov23}. 

Intuitively, the stars that are more loosely bound to the merging galaxy, i.e.~furthest from its center, are expected to be stripped during the first passages, and this is confirmed in simulations \citep[e.g.][]{Mori24}. As the merging galaxy passes through the Milky Way's potential well for the first time, it has high energy relative to the Milky Way, but with each passage it loses kinetic energy (Fig.~\ref{fig:am}, star symbols). Focusing only on the stars which now reside at Sun-like distances from the Galactic center (7-9\,kpc) shows qualitatively the same trend, see Fig.~\ref{fig:am}~(right). However, the most ($E>0\rm\,km^2\,s^{-2}$) and least ($E<-13\rm\,km^2\,s^{-2}$) energetic stars are missing, as they now typically reside at larger and smaller distances from the Galactic center, respectively.
From Fig.~\ref{fig:am} we conclude that the stars that were accreted first resided in the outer regions of Gaia Enceladus and will typically have higher orbital energy after joining the Milky Way. 

Fig.~\ref{fig:sel} shows the energy and angular momentum of the NS sample. The energy range, is in generally good agreement with simulations, see Fig.~\ref{fig:am} (right). By limiting ourselves to observations in the solar neighborhood, we do expect lesser range in both radius and energy. This qualitative result is robust, however, we caution that the quantitative scale might be affected by details in the scaling of the simulations \citep[e.g.][]{Pagnini23}.

To test the possibility of multiple passages of Gaia Enceladus, we want to explore the chemical signatures of stars potentially stripped at different passages, i.e.~stars with different energies. We separate the accreted stars into high- and low- energy subpopulations based on whether their orbital energies lie above or below the limit $E_{cut}=-4.5\cdot10^4$\,km$^{2}$\,s$^{-1}$. This is consistent with simulations, which show that stars with $E>-4.5\cdot10^4$\,km$^{2}$\,s$^{-1}$ are primarily expected to come from the outer regions of Gaia Enceladus, see Fig.~\ref{fig:am}. However, we do note that the quantitative scale of the kinetic energies is dependent on the adopted model for the Milky Way potential. Changing $E_{cut}$ within the refinements of our sample size does not affect our final conclusions, see Sec.~\ref{sec:ecut}.

\section{Elemental Abundance Trends} \label{sec:abu}

Using the definition from Fig.~\ref{fig:sel}, we separate the accreted low-$\alpha$ stars into high and low energy, with the aim of  investigating the chemical abundance trends of the two samples. 
It is well established that star formation and chemical enrichment is more efficient in the centers of galaxies compared to their outskirts, as seen by their ubiquitously observed metallicity gradients \citep[e.g.][]{Berg13,Tolstoy23,Fu24}. Therefore, if the stars with high energy truly come from the outer regions of Gaia Enceladus, they are expected to have experienced slower chemical enrichment compared to the inner regions traced by the low-energy stars, and this should be reflected in their chemical abundances.

\begin{figure}
\begin{center}
\includegraphics[width=0.95\hsize]{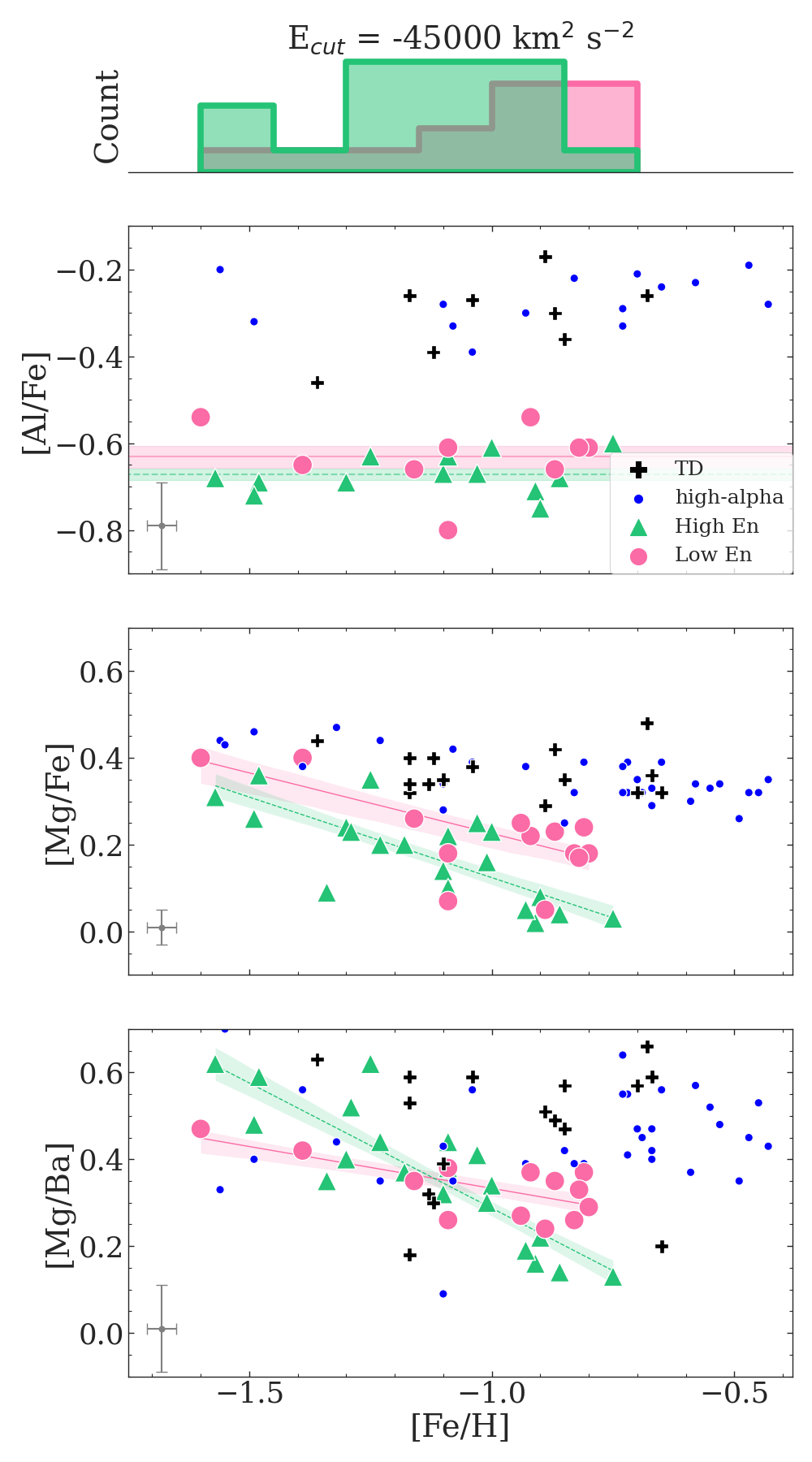} 
\caption{Chemical abundances for the accreted stellar sample with low (pink circles) and high (green triangles) energies, as defined in Fig.~\ref{fig:sel}. The average values for [Al/Fe] are shown with a line, while shaded area shows the error of the mean. For [Mg/Fe] and [Mg/Ba], trend lines are shown with 68\% confidence interval. Other abundance ratios are shown in Appendix~\ref{sec:xabu}. Top marginal plot shows the metallicity distribution of the accreted samples. For reference, the thick disk and high-$\alpha$ Milky Way populations created in situ are shown with small black crosses and blue circles, respectively. Representative error bar for individual measurements is shown in the bottom left corners. 
}
\label{fig:abu}
\end{center}
\end{figure}

First, we use our new Al measurements to refine the categorization of stars into the in-situ (high-$\alpha$) and accreted (low-$\alpha$) populations, which become hard to separate with $\alpha$-elements at low $\rm[Fe/H]<-1.2$, where the difference in [$\alpha$/Fe] is very small ($\lesssim0.1$\,dex), see Fig.~\ref{fig:abu} (middle panel). Based on our new data, we define a cut between the high- and low-$\alpha$ populations at $\rm[Al/Fe]=-0.5$, resulting in 9 stars with Al measurements in the low-energy sample and 13 with high energy. Our results show that the separation between the accreted and in-situ populations in [Al/Fe] is very clear ($\gtrsim0.3\,$dex), even at the lowest [Fe/H], see Fig.~\ref{fig:abu} (top panel). The new aluminum abundances reveal that one star at $\rm[Fe/H]=-1.4$ (named BD+07-4841, pink circle in Fig.~\ref{fig:abu}), which was previously categorized as high-$\alpha$ star, actually belongs to the accreted population (see also Ernandes et al. in prep.).

We can now compare the measured [Al/Fe] ratios in the high- and low-energy samples. The average aluminum for the accreted stars with high energy (as defined in Fig.~\ref{fig:sel}), is $\rm\langle[Al/Fe]\rangle_{\rm HE}=-0.67\pm0.01$ and for the accreted stars with low energy is $\rm \langle[Al/Fe]\rangle_{\rm LE}=-0.63\pm0.03$, using the error of the mean shown with shaded areas around the mean values in Fig.~\ref{fig:abu}. Performing a Kolmogorov–Smirnov (KS) test, the hypothesis that these two samples come from the same distribution is rejected, with $p<0.05$.
The two samples therefore show a statistically meaningful difference, with the high-energy stars having lower [Al/Fe] as expected for stars coming from regions of less efficient chemical enrichment. However, the difference of the average value is within 2$\sigma$, and we do note that aluminum is challenging to measure precisely, and the blue spectra needed for Al measurements were only available for a subset of the original NS sample (Sec.~\ref{sec:data}). 

More precise measurements can be made for Mg, and those exist for the entire NS sample of 33~accreted stars with known orbital energies (thereof 20 with high energy, and 13 low-energy stars). 
In Fig.~\ref{fig:abu} (middle panel) we see two evolutionary tracks of [Mg/Fe] as a function of [Fe/H]. The fits to these tracks were created by a linear regression and limited over the range where the data is available. The shaded areas represent the confidence interval of the linear regression estimates from bootstrap of 68\%. Again, the abundance ratios of stars with high energy reflect a birth environment experiencing a slightly less efficient chemical enrichment compared to the low-energy sample. 
As anticipated, the difference in [Mg/Fe] between the two samples is clearest at high $\rm[Fe/H]>-1$, where SNIa dominate the production of~Fe. The statistical significance of the [Mg/Fe] separation in the two subpopulations at high metallicity is corroborated with a KS test which gives $p<0.01$ for the samples at $\rm[Fe/H]>-1$, confirming that they have distinct distributions.
We emphasize that this separation is not only seen in Mg, but also in other $\alpha$-elements (see Appendix~\ref{sec:xabu}). 

\begin{figure*}
\begin{center}
\includegraphics[width=0.8\hsize]{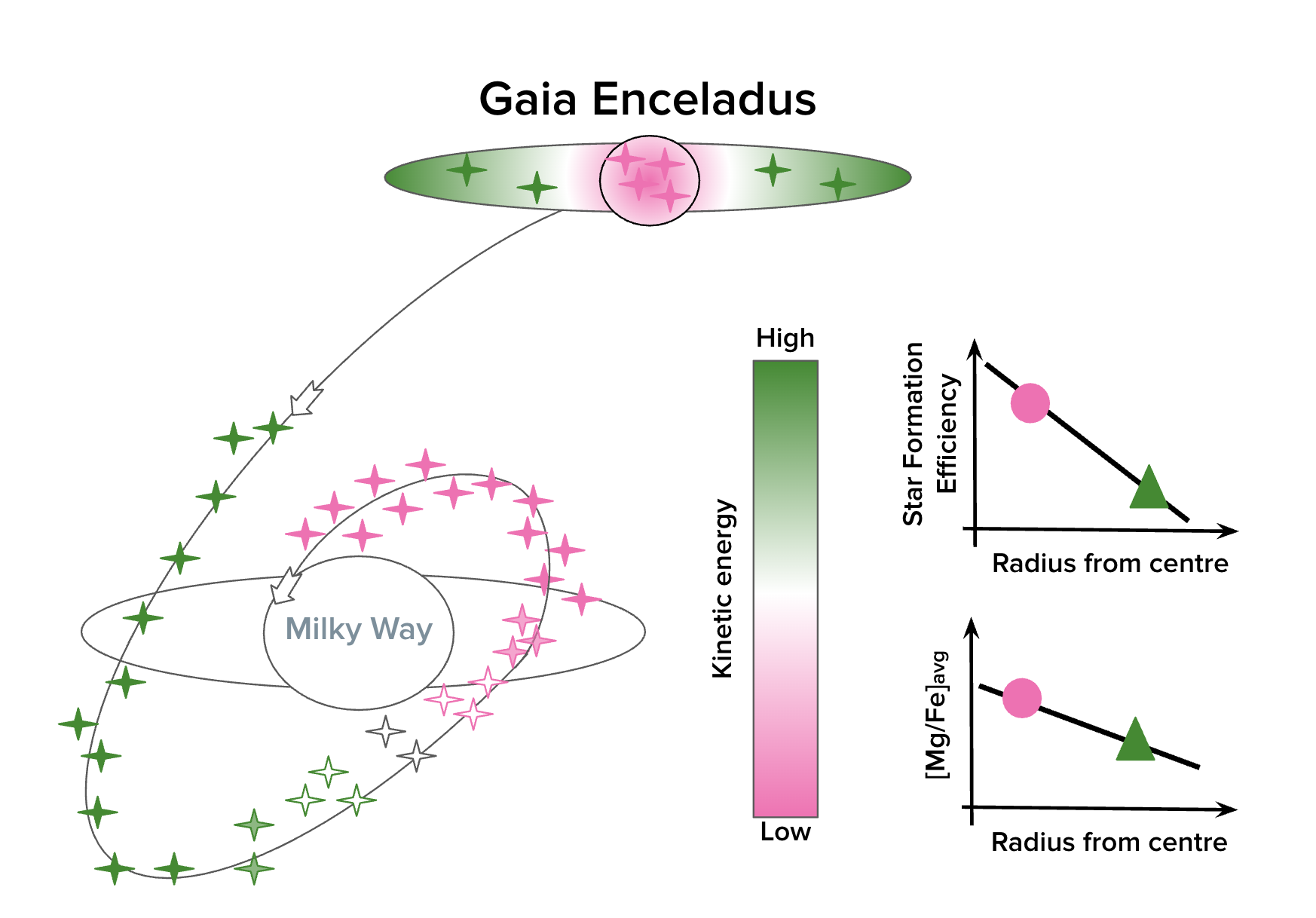} 
\caption{ Schematic figure showing how Gaia Enceladus fell into the potential of the Milky Way. First to be accreted with high kinetic energies are the outermost stars of Gaia Enceladus (green), from regions with less efficient star formation. These stars can subsequently be recognized by their lower [Mg/Fe] ratios at a given [Fe/H]. Conversely, the stars that originally resided closer to the center of Gaia Enceladus (pink) are accreted in later passages, thus having lower orbital energies, and the chemical signatures of more efficient star formation (e.g. higher [Mg/Fe] at high [Fe/H]). 
}
\label{fig:schematic}
\end{center}
\end{figure*} 

Finally, we show the [Mg/Ba] ratios in Fig.~\ref{fig:abu} (bottom panel). 
Yet again, we see that the abundance trend of the high-energy stars indicates that they come from a region with less efficient chemical enrichment compared to the low-energy sample. 
All stellar populations of different galactic origins are expected to have a similar value of $\rm[Mg/Fe]\approx+0.4$ at low $\rm [Fe/H]\lesssim-1.5$, since at early times Mg and Fe are co-produced in ccSN before the onset of SNIa. This is not the case for [Mg/Ba], because Mg and Ba have distinctly different production sites. The [Mg/Ba] ratio at low [Fe/H] can therefore vary between galaxies and different regions within galaxies \citep[e.g.][]{Skuladottir20}. In Fig.~\ref{fig:abu}, we notice that the [Mg/Ba] slope of the high-energy sample is significantly steeper than for low-energy stars. This is expected in a region of less efficient star formation, as more intermediate-mass stars reach their AGB phase in the time taken for the region to enrich to the same [Fe/H]. Performing a two-dimensional KS test is beyond the scope of this work, but comparing [Mg/Ba] in the full samples is misleading as there is clearly a trend with [Fe/H], which is not fully captured with a one-dimensional test. If we instead perform a focused KS test at high $\rm[Fe/H]>-1$ we get $p<0.01$, confirming that [Mg/Ba] separates in the two subpopulations in a statistically meaningful way at high [Fe/H]. 

We note that in the bottom panel of Fig.~\ref{fig:abu} we have excluded one Ba-rich star ($\rm[Mg/Ba]<0$) with low kinetic energy. High Ba abundances are typically obtained through binary transfer from an AGB stellar companion \citep[e.g.][]{Karakas14}, and such stars are therefore not representative of the chemical evolution history of their host galaxy. 

In summary, the chemical abundance ratios in Fig.~\ref{fig:abu} consistently indicate that the high-energy accreted sample was formed in a region of less efficient chemical enrichment compared to the low-energy accreted stars. This is further supported by the metallicity distribution of the sample (Fig.~\ref{fig:abu}, top marginal plot), where the low energy stars are on average more metal-rich, $\rm \langle[Fe/H]\rangle_{LE}=-1.03\pm0.06$, compared to the high energy sample $\rm \langle[Fe/H]\rangle_{HE}=-1.16\pm0.05$. A one-dimensional KS test of [Fe/H] for the entire low- and high-energy samples gives a p-value of $p=0.19$. Together the metallicity and abundance ratios therefore show that the sample with high energy comes from a region of less efficient star formation, compared to the low-energy stars.

\section{Discussion and Conclusions} \label{sec:con}

In this paper, we investigate the hypothesis that evidence of multiple passages of the Gaia Enceladus galaxy can be seen in the high-quality kinematics and chemical abundances of accreted stars in the Galactic halo. 
Guided by simulations, we divided the accreted stars from the historical sample of \citet{Nissen10}, into two groups, with high and low kinetic energies, as separated by the limit $E_{cut}=-4.5\cdot10^4$\,km$^{2}$\,s$^{-1}$ (see Fig.~\ref{fig:sel}). Accreted stars with high energy are predicted to be stripped from the outskirts of Gaia Enceladus during the early passages, while stars which currently have low kinematic energy should come from the more inner regions during later passages (Sec.~\ref{sec:energy}).

For the first time we provide aluminum measurements for this sample, showing that [Al/Fe] ratios are the best tracers of accreted stars with $\rm[Fe/H]<-1.2$ (see also Ernandes et al. in prep.). We investigated the abundance ratios of [Al/Fe], [Mg/Fe], and [Mg/Ba], which are known to separate in systems with different star formation histories (Sec.~\ref{sec:abu}). In all observed abundance ratios the stars with high energy show clear evidence of coming from an environment which experienced less efficient chemical enrichment compared to those with low energy. This was confirmed with the low energy sample having on average higher metallicity, $\rm \langle[Fe/H]\rangle_{LE}>\langle[Fe/H]\rangle_{HE}$.

Both kinematically and chemically, the high-energy accreted sample is consistent with what the community has defined as Gaia Enceladus \citep[e.g.][]{Feuillet21,Ernandes24,Davies24}. But what is the origin of the low-energy subpopulation? Previously, these stars have been proposed to belong to Thamnos \citep{Nissen24} or Eos \citep{Matsuno24}. However, the Thamnos merger event is incompatible with the kinematics of the low-energy population as pointed out by \citet{Matsuno24}. Based on the low $\rm[Al/Fe]<-0.5$ values measured here in this sample, the in-situ formed Eos can now also be excluded, since one of its main characteristics are high [Al/Fe] abundance ratios \citep{Myeong22}.
 
More generally, the hypothesis of the high- and low-energy populations belonging to two separate merger events can be rejected for several reasons: 1)~For a second merger to appear unambiguously in such a small sample ($\approx$35\, accreted stars), would require it to be of comparable mass to Gaia Enceladus, and the higher metallicity of the low-energy population suggests an even larger merger. No evidence of such a merger has been found. 2)~It is unphysical to expect that a major merger of this size would not experience several passages
(resulting in a higher-energy counterpart, whose extent may depend on
the time of accretion). This goes against theoretical predictions \citep[e.g.][]{Khoperskov23a,Khoperskov23b,Mori24}, and is contrary to what is seen observationally, e.g.~in the case of the Sagittarius dwarf spheroidal galaxy currently being disrupted by the Milky Way \citep{Johnston95}. 3)~It is highly improbable that two separate galaxies independently experienced chemical enrichment that was so extremely similar to each other, as seen by the very close abundance trends of accreted stars with high and low energy in Fig.~\ref{fig:abu}. Such close similarities are not seen in observations of the different satellite galaxies of the Milky Way, where the typical abundance differences in [Mg/Fe] at the highest [Fe/H] reach $\gtrsim0.3$\,dex \citep[e.g.][]{Tolstoy09}. Furthermore, no two satellite galaxies are known to have indistinguishable [Ca/Fe] as is the case for the high- and low-energy populations (see Appendix~\ref{sec:xabu}). However, comparable differences (as shown in Fig.~\ref{fig:abu}) are seen in the inner and outer regions of the Sculptor dwarf spheriodal galaxy (Lucchesi et al. in prep.). 

Therefore, we conclude that these two subpopulations of low and high energy both belong to Gaia Enceladus, and that they correspond to different accretion epochs (Fig.~\ref{fig:schematic}). In agreement both with simulations and observations, the stars with high energy were accreted earlier, during an initial passage, and stripped from the outer regions of Gaia Enceladus. All the available abundance information (Fig.~\ref{fig:abu}) points to a very similar chemical enrichment history between the two groups, where the stars with high energy formed in a region experiencing less efficient star formation, as expected when comparing the outer region of a galaxy to its inner parts. 

Furthermore, the accreted stars with the highest observed $L_{z}>0.5\cdot10^{3}\rm\,km\,s^{-1}\,kpc$, which are predicted to be stars initially residing at large distances from the center of Gaia Enceladus (see Fig.~\ref{fig:am}), all have low $\rm[Fe/H]<-1$, consistent with coming from the outermost least gravitationally bound and least chemically evolved regions of the galaxy. 

Unfortunately, our small sample size limits us from further exploring whether the abundance trends change gradually with increasing energy. Furthermore, we are not able to assign stars to specific passages, e.g.~we cannot determine whether the high-energy sample is dominated by the first passage, or if it is a combination of the first and second passages (see Fig.~\ref{fig:am}). Simulations predict over-densities in the $E-L_z$ plane corresponding to specific passages \citep{Mori24}, which should be observable with larger sample sizes, making the specific stripping passage of stars easier to identify. We emphasize that for the chemical abundance analysis of such a sample, very high quality spectrum is preferred ($SNR>100$, $R\gtrsim40\,000$) as is having a small range in stellar parameters, since a large range in $\log g$ and/or $T_{\rm eff}$ can result in very different non-LTE effects (see e.g.~\citealt{Koutsouridou25}), causing a measurement scatter that could hide or distort intrinsic trends. Furthermore, having more groups providing simulations of the Gaia Enceladus merger would help to establish the robustness of the theoretical predictions.

For the first time, we have presented observational evidence for Gaia Enceladus experiencing at least two passages before being dispersed into the Milky Way. This discovery represents a new phase in the investigation of the hierarchical galaxy formation history of our Milky Way. Instead of looking at a merger as a singular event, we are able to identify several passages, characterized by individual stars. 
Our results indicate that the general properties of Gaia Enceladus (and perhaps other merger events) need to be revisited: by defining the galaxy only by its outskirts (high-energy stars) while neglecting the inner parts (low-energy stars), both the metallicity and stellar mass could be underestimated, as could the timescale of how long this merger event lasted. With the next generation of data and theoretical models, this discovery gives us the unique opportunity to study in extraordinary detail the spatially resolved star formation history and spatially resolved chemical enrichment history of a galaxy that died $\sim$10~Gyr ago. Such high-precision studies can drastically improve our knowledge of galaxy mergers in general, creating invaluable synergies with the more numerous observations of extra-Galactic mergers at higher redshifts, bringing us significantly closer to understanding the complexities of hierarchical galaxy formation throughout the history of our Universe.

\begin{acknowledgments}
This project has received funding from the European Research Council Executive Agency (ERCEA) under the European Union's Horizon Europe research and innovation program (acronym TREASURES, grant agreement No 101117455, PI Skúladóttir). H.E., S.F. and D.F. were supported by a project grant from the Knut and Alice Wallenberg Foundation (KAW 2020.0061 Galactic Time Machine, PI Feltzing). This project was supported by funds from the Crafoord foundation (reference 20230890) and from the Royal Physiographic Society of Lund. H.E. was supported by a grant from the Royal Swedish Academy of Sciences and the Royal Physiographic Society of Lund. D.F. acknowledges funding from the
Swedish Research Council grant 2022-03274. This work has made use of data from the European Space Agency (ESA) mission, processed by the {\it Gaia}
Data Processing and Analysis Consortium (DPAC). Funding for the DPAC
has been provided by national institutions, in particular the institutions
participating in the {\it Gaia} Multilateral Agreement. Views and opinions expressed are however those of the author(s) only and do not necessarily reflect those of the ERC or other granting authority. Neither the European Union nor other granting authority can be held responsible for them. H.E. would like to thank Kaio Zampronho Batista for helping with color determinations for the plots.
\end{acknowledgments}

\begin{contribution}
The data reduction was done by Á.S., who was also the PI of the observational proposals used in this work. The chemical abundance analysis was led by H.E. and done in collaboration with Á.S, with the assistance of R.L and S.F. The kinetic energies were calculated by D.F., who also played a lead role in identifying the two chemically distinct substructures of the sample, while S.F., Á.S. and H.E. participated in the process. A.M. and P.D.M are responsible for the simulations displayed here. H.E. created the figures (other than Fig.~\ref{fig:am}, done by A.M.), with input from other co-authors. The first author took the lead in structuring the idea and writing the paper, but all co-authors contributed substantially to the process. 
\end{contribution}

\facilities{ESO VLT/UVES, Gaia DR3}

\software{ESO reduction pipeline \citep{Freudling13}, Marcs models \citep{Gustafsson08}, TURBOSPECTRUM \citep{Plez12}
          }

\bibliography{gse}{}

\begin{thebibliography}{}
\expandafter\ifx\csname natexlab\endcsname\relax\def\natexlab#1{#1}\fi
\providecommand{\url}[1]{\href{#1}{#1}}
\providecommand{\dodoi}[1]{doi:~\href{http://doi.org/#1}{\nolinkurl{#1}}}
\providecommand{\doeprint}[1]{\href{http://ascl.net/#1}{\nolinkurl{http://ascl.net/#1}}}
\providecommand{\doarXiv}[1]{\href{https://arxiv.org/abs/#1}{\nolinkurl{https://arxiv.org/abs/#1}}}

\bibitem[{ {Abdurro'uf} {et~al.}(2022){Abdurro'uf}, {Accetta}, {Aerts}, {Silva
  Aguirre}, {Ahumada}, {Ajgaonkar}, {Filiz Ak}, {Alam}, {Allende Prieto},
  {Almeida}, {Anders}, {Anderson}, {Andrews}, {Anguiano}, {Aquino-Ort{\'\i}z},
  {Arag{\'o}n-Salamanca}, {Argudo-Fern{\'a}ndez}, {Ata}, {Aubert},
  {Avila-Reese}, {Badenes}, {Barb{\'a}}, {Barger}, {Barrera-Ballesteros},
  {Beaton}, {Beers}, {Belfiore}, {Bender}, {Bernardi}, {Bershady}, {Beutler},
  {Bidin}, {Bird}, {Bizyaev}, {Blanc}, {Blanton}, {Boardman}, {Bolton},
  {Boquien}, {Borissova}, {Bovy}, {Brandt}, {Brown}, {Brownstein}, {Brusa},
  {Buchner}, {Bundy}, {Burchett}, {Bureau}, {Burgasser}, {Cabang}, {Campbell},
  {Cappellari}, {Carlberg}, {Wanderley}, {Carrera}, {Cash}, {Chen}, {Chen},
  {Cherinka}, {Chiappini}, {Choi}, {Chojnowski}, {Chung}, {Clerc}, {Cohen},
  {Comerford}, {Comparat}, {da Costa}, {Covey}, {Crane}, {Cruz-Gonzalez},
  {Culhane}, {Cunha}, {Dai}, {Damke}, {Darling}, {Davidson}, {Davies},
  {Dawson}, {De Lee}, {Diamond-Stanic}, {Cano-D{\'\i}az}, {S{\'a}nchez},
  {Donor}, {Duckworth}, {Dwelly}, {Eisenstein}, {Elsworth}, {Emsellem},
  {Eracleous}, {Escoffier}, {Fan}, {Farr}, {Feng}, {Fern{\'a}ndez-Trincado},
  {Feuillet}, {Filipp}, {Fillingham}, {Frinchaboy}, {Fromenteau}, {Galbany},
  {Garc{\'\i}a}, {Garc{\'\i}a-Hern{\'a}ndez}, {Ge}, {Geisler}, {Gelfand},
  {G{\'e}ron}, {Gibson}, {Goddy}, {Godoy-Rivera}, {Grabowski}, {Green},
  {Greener}, {Grier}, {Griffith}, {Guo}, {Guy}, {Hadjara}, {Harding},
  {Hasselquist}, {Hayes}, {Hearty}, {Hern{\'a}ndez}, {Hill}, {Hogg},
  {Holtzman}, {Horta}, {Hsieh}, {Hsu}, {Hsu}, {Huber}, {Huertas-Company},
  {Hutchinson}, {Hwang}, {Ibarra-Medel}, {Chitham}, {Ilha}, {Imig}, {Jaekle},
  {Jayasinghe}, {Ji}, {Johnson}, {Jones}, {J{\"o}nsson}, {Katkov}, {Khalatyan},
  {Kinemuchi}, {Kisku}, {Knapen}, {Kneib}, {Kollmeier}, {Kong}, {Kounkel},
  {Kreckel}, {Krishnarao}, {Lacerna}, {Lane}, {Langgin}, {Lavender}, {Law},
  {Lazarz}, {Leung}, {Leung}, {Lewis}, {Li}, {Li}, {Lian}, {Liang}, {Lin},
  {Lin}, {Lin}, {Lintott}, {Long}, {Longa-Pe{\~n}a}, {L{\'o}pez-Cob{\'a}},
  {Lu}, {Lundgren}, {Luo}, {Mackereth}, {de la Macorra}, {Mahadevan},
  {Majewski}, {Manchado}, {Mandeville}, {Maraston}, {Margalef-Bentabol},
  {Masseron}, {Masters}, {Mathur}, {McDermid}, {Mckay}, {Merloni},
  {Merrifield}, {Meszaros}, {Miglio}, {Di Mille}, {Minniti}, {Minsley}, \&
  {Monachesi}}]{Abdurrouf22}
{Abdurro'uf}, {Accetta}, K., {Aerts}, C., {et~al.} 2022, \bibinfo{title}{{The
  Seventeenth Data Release of the Sloan Digital Sky Surveys: Complete Release
  of MaNGA, MaStar, and APOGEE-2 Data},} \apjs, 259, 35,
  \dodoi{10.3847/1538-4365/ac4414}

\bibitem[{R. {Alvarez} \& B. {Plez}(1998){Alvarez} \& {Plez}}]{Alvarez98}
{Alvarez}, R., \& {Plez}, B. 1998, \bibinfo{title}{{Near-infrared narrow-band
  photometry of M-giant and Mira stars: models meet observations},} \aap, 330,
  1109, \dodoi{10.48550/arXiv.astro-ph/9710157}

\bibitem[{J.~A.~S. {Amarante} {et~al.}(2022){Amarante}, {Debattista}, {Beraldo
  e Silva}, {Laporte}, \& {Deg}}]{Amarante22}
{Amarante}, J. A.~S., {Debattista}, V.~P., {Beraldo e Silva}, L., {Laporte}, C.
  F.~P., \& {Deg}, N. 2022, \bibinfo{title}{{Gastro Library. I. The Simulated
  Chemodynamical Properties of Several Gaia-Sausage-Enceladus-like Stellar
  Halos},} \apj, 937, 12, \dodoi{10.3847/1538-4357/ac8b0d}

\bibitem[{V. {Belokurov} {et~al.}(2018){Belokurov}, {Erkal}, {Evans},
  {Koposov}, \& {Deason}}]{Belukorov18}
{Belokurov}, V., {Erkal}, D., {Evans}, N.~W., {Koposov}, S.~E., \& {Deason},
  A.~J. 2018, \bibinfo{title}{{Co-formation of the disc and the stellar halo},}
  \mnras, 478, 611, \dodoi{10.1093/mnras/sty982}

\bibitem[{V. {Belokurov} {et~al.}(2023){Belokurov}, {Vasiliev}, {Deason},
  {Koposov}, {Fattahi}, {Dillamore}, {Davies}, \& {Grand}}]{Belokurov23}
{Belokurov}, V., {Vasiliev}, E., {Deason}, A.~J., {et~al.} 2023,
  \bibinfo{title}{{Energy wrinkles and phase-space folds of the last major
  merger},} \mnras, 518, 6200, \dodoi{10.1093/mnras/stac3436}

\bibitem[{D.~A. {Berg} {et~al.}(2013){Berg}, {Skillman}, {Garnett}, {Croxall},
  {Marble}, {Smith}, {Gordon}, \& {Kennicutt}}]{Berg13}
{Berg}, D.~A., {Skillman}, E.~D., {Garnett}, D.~R., {et~al.} 2013,
  \bibinfo{title}{{New Radial Abundance Gradients for NGC 628 and NGC 2403},}
  \apj, 775, 128, \dodoi{10.1088/0004-637X/775/2/128}

\bibitem[{J. {Binney}(2012){Binney}}]{Binney2012}
{Binney}, J. 2012, \bibinfo{title}{{Actions for axisymmetric potentials},}
  \mnras, 426, 1324, \dodoi{10.1111/j.1365-2966.2012.21757.x}

\bibitem[{J. {Bovy}(2015){Bovy}}]{Bovy15}
{Bovy}, J. 2015, \bibinfo{title}{{galpy: A python Library for Galactic
  Dynamics},} \apjs, 216, 29, \dodoi{10.1088/0067-0049/216/2/29}

\bibitem[{J. {Bovy} \& H.-W. {Rix}(2013){Bovy} \& {Rix}}]{Bovy2013}
{Bovy}, J., \& {Rix}, H.-W. 2013, \bibinfo{title}{{A Direct Dynamical
  Measurement of the Milky Way's Disk Surface Density Profile, Disk Scale
  Length, and Dark Matter Profile at 4 kpc <\raisebox{-0.5ex}\textasciitilde R
  <\raisebox{-0.5ex}\textasciitilde 9 kpc},} \apj, 779, 115,
  \dodoi{10.1088/0004-637X/779/2/115}

\bibitem[{P. {Das} {et~al.}(2020){Das}, {Hawkins}, \& {Jofr{\'e}}}]{Das2020}
{Das}, P., {Hawkins}, K., \& {Jofr{\'e}}, P. 2020, \bibinfo{title}{{Ages and
  kinematics of chemically selected, accreted Milky Way halo stars},} \mnras,
  493, 5195, \dodoi{10.1093/mnras/stz3537}

\bibitem[{E.~Y. {Davies} {et~al.}(2024){Davies}, {Belokurov}, {Kravtsov},
  {Monty}, {Myeong}, {Evans}, \& {Kane}}]{Davies24}
{Davies}, E.~Y., {Belokurov}, V., {Kravtsov}, A., {et~al.} 2024,
  \bibinfo{title}{{Blind source separation of the stellar halo},} arXiv
  e-prints, arXiv:2410.21365, \dodoi{10.48550/arXiv.2410.21365}

\bibitem[{P. {Di Matteo}(2016){Di Matteo}}]{DiMatteo16}
{Di Matteo}, P. 2016, \bibinfo{title}{{The Disc Origin of the Milky Way
  Bulge},} \pasa, 33, e027, \dodoi{10.1017/pasa.2016.11}

\bibitem[{H. {Ernandes} {et~al.}(2024){Ernandes}, {Feuillet}, {Feltzing}, \&
  {Sk{\'u}lad{\'o}ttir}}]{Ernandes24}
{Ernandes}, H., {Feuillet}, D., {Feltzing}, S., \& {Sk{\'u}lad{\'o}ttir},
  {\'A}. 2024, \bibinfo{title}{{Gaia-Sausage-Enceladus star formation history
  as revealed by detailed elemental abundances: An archival study using SAGA
  data},} \aap, 691, A333, \dodoi{10.1051/0004-6361/202450827}

\bibitem[{S. {Feltzing} \& D. {Feuillet}(2023){Feltzing} \&
  {Feuillet}}]{Feltzing23}
{Feltzing}, S., \& {Feuillet}, D. 2023, \bibinfo{title}{{The Metal-weak Milky
  Way Stellar Disk Hidden in the Gaia-Sausage-Enceladus Debris: The APOGEE DR17
  View},} \apj, 953, 143, \dodoi{10.3847/1538-4357/ace185}

\bibitem[{D.~K. {Feuillet} {et~al.}(2020){Feuillet}, {Feltzing}, {Sahlholdt},
  \& {Casagrande}}]{Feuillet20}
{Feuillet}, D.~K., {Feltzing}, S., {Sahlholdt}, C.~L., \& {Casagrande}, L.
  2020, \bibinfo{title}{{The SkyMapper-Gaia RVS view of the
  Gaia-Enceladus-Sausage - an investigation of the metallicity and mass of the
  Milky Way's last major merger},} \mnras, 497, 109,
  \dodoi{10.1093/mnras/staa1888}

\bibitem[{D.~K. {Feuillet} {et~al.}(2021){Feuillet}, {Sahlholdt}, {Feltzing},
  \& {Casagrande}}]{Feuillet21}
{Feuillet}, D.~K., {Sahlholdt}, C.~L., {Feltzing}, S., \& {Casagrande}, L.
  2021, \bibinfo{title}{{Selecting accreted populations: metallicity, elemental
  abundances, and ages of the Gaia-Sausage-Enceladus and Sequoia populations},}
  \mnras, 508, 1489, \dodoi{10.1093/mnras/stab2614}

\bibitem[{W. {Freudling} {et~al.}(2013){Freudling}, {Romaniello}, {Bramich},
  {Ballester}, {Forchi}, {Garc{\'{\i}}a-Dabl{\'o}}, {Moehler}, \&
  {Neeser}}]{Freudling13}
{Freudling}, W., {Romaniello}, M., {Bramich}, D.~M., {et~al.} 2013,
  \bibinfo{title}{{Automated data reduction workflows for astronomy. The ESO
  Reflex environment},} \aap, 559, A96, \dodoi{10.1051/0004-6361/201322494}

\bibitem[{S.~W. {Fu} {et~al.}(2024){Fu}, {Weisz}, {Starkenburg}, {Martin},
  {Collins}, {Savino}, {Boylan-Kolchin}, {C{\^o}t{\'e}}, {Dolphin}, {Longeard},
  {Mateo}, {Mercado}, {Sandford}, \& {Skillman}}]{Fu24}
{Fu}, S.~W., {Weisz}, D.~R., {Starkenburg}, E., {et~al.} 2024,
  \bibinfo{title}{{Stellar Metallicities and Gradients in the Faint M31
  Satellites Andromeda XVI and Andromeda XXVIII},} \apj, 975, 2,
  \dodoi{10.3847/1538-4357/ad76a2}

\bibitem[{ {Gaia Collaboration} {et~al.}(2016){Gaia Collaboration}, {Prusti},
  {de Bruijne}, {Brown}, {Vallenari}, {Babusiaux}, {Bailer-Jones}, {Bastian},
  {Biermann}, {Evans}, {Eyer}, {Jansen}, {Jordi}, {Klioner}, {Lammers},
  {Lindegren}, {Luri}, {Mignard}, {Milligan}, {Panem}, {Poinsignon},
  {Pourbaix}, {Randich}, {Sarri}, {Sartoretti}, {Siddiqui}, {Soubiran},
  {Valette}, {van Leeuwen}, {Walton}, {Aerts}, {Arenou}, {Cropper}, {Drimmel},
  {H{\o}g}, {Katz}, {Lattanzi}, {O'Mullane}, {Grebel}, {Holland}, {Huc},
  {Passot}, {Bramante}, {Cacciari}, {Casta{\~n}eda}, {Chaoul}, {Cheek}, {De
  Angeli}, {Fabricius}, {Guerra}, {Hern{\'a}ndez}, {Jean-Antoine-Piccolo},
  {Masana}, {Messineo}, {Mowlavi}, {Nienartowicz}, {Ord{\'o}{\~n}ez-Blanco},
  {Panuzzo}, {Portell}, {Richards}, {Riello}, {Seabroke}, {Tanga},
  {Th{\'e}venin}, {Torra}, {Els}, {Gracia-Abril}, {Comoretto},
  {Garcia-Reinaldos}, {Lock}, {Mercier}, {Altmann}, {Andrae}, {Astraatmadja},
  {Bellas-Velidis}, {Benson}, {Berthier}, {Blomme}, {Busso}, {Carry},
  {Cellino}, {Clementini}, {Cowell}, {Creevey}, {Cuypers}, {Davidson}, {De
  Ridder}, {de Torres}, {Delchambre}, {Dell'Oro}, {Ducourant}, {Fr{\'e}mat},
  {Garc{\'\i}a-Torres}, {Gosset}, {Halbwachs}, {Hambly}, {Harrison}, {Hauser},
  {Hestroffer}, {Hodgkin}, {Huckle}, {Hutton}, {Jasniewicz}, {Jordan},
  {Kontizas}, {Korn}, {Lanzafame}, {Manteiga}, {Moitinho}, {Muinonen},
  {Osinde}, {Pancino}, {Pauwels}, {Petit}, {Recio-Blanco}, {Robin}, {Sarro},
  {Siopis}, {Smith}, {Smith}, {Sozzetti}, {Thuillot}, {van Reeven}, {Viala},
  {Abbas}, {Abreu Aramburu}, {Accart}, {Aguado}, {Allan}, {Allasia},
  {Altavilla}, {{\'A}lvarez}, {Alves}, {Anderson}, {Andrei}, {Anglada Varela},
  {Antiche}, {Antoja}, {Ant{\'o}n}, {Arcay}, {Atzei}, {Ayache}, {Bach},
  {Baker}, {Balaguer-N{\'u}{\~n}ez}, {Barache}, {Barata}, {Barbier}, {Barblan},
  {Baroni}, {Barrado y Navascu{\'e}s}, {Barros}, {Barstow}, {Becciani},
  {Bellazzini}, {Bellei}, {Bello Garc{\'\i}a}, {Belokurov}, {Bendjoya},
  {Berihuete}, {Bianchi}, {Bienaym{\'e}}, {Billebaud}, {Blagorodnova},
  {Blanco-Cuaresma}, {Boch}, {Bombrun}, {Borrachero}, {Bouquillon}, {Bourda},
  {Bouy}, {Bragaglia}, {Breddels}, {Brouillet}, {Br{\"u}semeister},
  {Bucciarelli}, {Budnik}, {Burgess}, {Burgon}, {Burlacu}, {Busonero}, {Buzzi},
  {Caffau}, {Cambras}, {Campbell}, {Cancelliere}, {Cantat-Gaudin}, {Carlucci},
  {Carrasco}, {Castellani}, {Charlot}, {Charnas}, {Charvet}, {Chassat},
  {Chiavassa}, {Clotet}, {Cocozza}, {Collins}, {Collins}, \&
  {Costigan}}]{Gaia16}
{Gaia Collaboration}, {Prusti}, T., {de Bruijne}, J.~H.~J., {et~al.} 2016,
  \bibinfo{title}{{The Gaia mission},} \aap, 595, A1,
  \dodoi{10.1051/0004-6361/201629272}

\bibitem[{ {Gaia Collaboration} {et~al.}(2023){Gaia Collaboration},
  {Vallenari}, {Brown}, {Prusti}, {de Bruijne}, {Arenou}, {Babusiaux},
  {Biermann}, {Creevey}, {Ducourant}, {Evans}, {Eyer}, {Guerra}, {Hutton},
  {Jordi}, {Klioner}, {Lammers}, {Lindegren}, {Luri}, {Mignard}, {Panem},
  {Pourbaix}, {Randich}, {Sartoretti}, {Soubiran}, {Tanga}, {Walton},
  {Bailer-Jones}, {Bastian}, {Drimmel}, {Jansen}, {Katz}, {Lattanzi}, {van
  Leeuwen}, {Bakker}, {Cacciari}, {Casta{\~n}eda}, {De Angeli}, {Fabricius},
  {Fouesneau}, {Fr{\'e}mat}, {Galluccio}, {Guerrier}, {Heiter}, {Masana},
  {Messineo}, {Mowlavi}, {Nicolas}, {Nienartowicz}, {Pailler}, {Panuzzo},
  {Riclet}, {Roux}, {Seabroke}, {Sordo}, {Th{\'e}venin}, {Gracia-Abril},
  {Portell}, {Teyssier}, {Altmann}, {Andrae}, {Audard}, {Bellas-Velidis},
  {Benson}, {Berthier}, {Blomme}, {Burgess}, {Busonero}, {Busso},
  {C{\'a}novas}, {Carry}, {Cellino}, {Cheek}, {Clementini}, {Damerdji},
  {Davidson}, {de Teodoro}, {Nu{\~n}ez Campos}, {Delchambre}, {Dell'Oro},
  {Esquej}, {Fern{\'a}ndez-Hern{\'a}ndez}, {Fraile}, {Garabato},
  {Garc{\'\i}a-Lario}, {Gosset}, {Haigron}, {Halbwachs}, {Hambly}, {Harrison},
  {Hern{\'a}ndez}, {Hestroffer}, {Hodgkin}, {Holl}, {Jan{\ss}en}, {Jevardat de
  Fombelle}, {Jordan}, {Krone-Martins}, {Lanzafame}, {L{\"o}ffler}, {Marchal},
  {Marrese}, {Moitinho}, {Muinonen}, {Osborne}, {Pancino}, {Pauwels},
  {Recio-Blanco}, {Reyl{\'e}}, {Riello}, {Rimoldini}, {Roegiers}, {Rybizki},
  {Sarro}, {Siopis}, {Smith}, {Sozzetti}, {Utrilla}, {van Leeuwen}, {Abbas},
  {{\'A}brah{\'a}m}, {Abreu Aramburu}, {Aerts}, {Aguado}, {Ajaj},
  {Aldea-Montero}, {Altavilla}, {{\'A}lvarez}, {Alves}, {Anders}, {Anderson},
  {Anglada Varela}, {Antoja}, {Baines}, {Baker}, {Balaguer-N{\'u}{\~n}ez},
  {Balbinot}, {Balog}, {Barache}, {Barbato}, {Barros}, {Barstow},
  {Bartolom{\'e}}, {Bassilana}, {Bauchet}, {Becciani}, {Bellazzini},
  {Berihuete}, {Bernet}, {Bertone}, {Bianchi}, {Binnenfeld}, {Blanco-Cuaresma},
  {Blazere}, {Boch}, {Bombrun}, {Bossini}, {Bouquillon}, {Bragaglia},
  {Bramante}, {Breedt}, {Bressan}, {Brouillet}, {Brugaletta}, {Bucciarelli},
  {Burlacu}, {Butkevich}, {Buzzi}, {Caffau}, {Cancelliere}, {Cantat-Gaudin},
  {Carballo}, {Carlucci}, {Carnerero}, {Carrasco}, {Casamiquela}, {Castellani},
  {Castro-Ginard}, {Chaoul}, {Charlot}, {Chemin}, {Chiaramida}, {Chiavassa},
  {Chornay}, {Comoretto}, {Contursi}, {Cooper}, {Cornez}, {Cowell}, {Crifo},
  {Cropper}, {Crosta}, {Crowley}, {Dafonte}, {Dapergolas}, {David}, {David},
  {de Laverny}, {De Luise}, \& {De March}}]{Gaia23}
{Gaia Collaboration}, {Vallenari}, A., {Brown}, A.~G.~A., {et~al.} 2023,
  \bibinfo{title}{{Gaia Data Release 3. Summary of the content and survey
  properties},} \aap, 674, A1, \dodoi{10.1051/0004-6361/202243940}

\bibitem[{J.~M. {Gerber} {et~al.}(2023){Gerber}, {Magg}, {Plez}, {Bergemann},
  {Heiter}, {Olander}, \& {Hoppe}}]{Gerber23}
{Gerber}, J.~M., {Magg}, E., {Plez}, B., {et~al.} 2023,
  \bibinfo{title}{{Non-LTE radiative transfer with Turbospectrum},} \aap, 669,
  A43, \dodoi{10.1051/0004-6361/202243673}

\bibitem[{B. {Gustafsson} {et~al.}(2008){Gustafsson}, {Edvardsson}, {Eriksson},
  {J{\o}rgensen}, {Nordlund}, \& {Plez}}]{Gustafsson08}
{Gustafsson}, B., {Edvardsson}, B., {Eriksson}, K., {et~al.} 2008,
  \bibinfo{title}{{A grid of MARCS model atmospheres for late-type stars. I.
  Methods and general properties},} \aap, 486, 951,
  \dodoi{10.1051/0004-6361:200809724}

\bibitem[{K. {Hawkins} {et~al.}(2015){Hawkins}, {Jofr{\'e}}, {Masseron}, \&
  {Gilmore}}]{Hawkins2015}
{Hawkins}, K., {Jofr{\'e}}, P., {Masseron}, T., \& {Gilmore}, G. 2015,
  \bibinfo{title}{{Using chemical tagging to redefine the interface of the
  Galactic disc and halo},} \mnras, 453, 758, \dodoi{10.1093/mnras/stv1586}

\bibitem[{M.~R. {Hayden} {et~al.}(2015){Hayden}, {Bovy}, {Holtzman}, {Nidever},
  {Bird}, {Weinberg}, {Andrews}, {Majewski}, {Allende Prieto}, {Anders},
  {Beers}, {Bizyaev}, {Chiappini}, {Cunha}, {Frinchaboy},
  {Garc{\'\i}a-Her{\'n}andez}, {Garc{\'\i}a P{\'e}rez}, {Girardi}, {Harding},
  {Hearty}, {Johnson}, {M{\'e}sz{\'a}ros}, {Minchev}, {O'Connell}, {Pan},
  {Robin}, {Schiavon}, {Schneider}, {Schultheis}, {Shetrone}, {Skrutskie},
  {Steinmetz}, {Smith}, {Wilson}, {Zamora}, \& {Zasowski}}]{Hayden15}
{Hayden}, M.~R., {Bovy}, J., {Holtzman}, J.~A., {et~al.} 2015,
  \bibinfo{title}{{Chemical Cartography with APOGEE: Metallicity Distribution
  Functions and the Chemical Structure of the Milky Way Disk},} \apj, 808, 132,
  \dodoi{10.1088/0004-637X/808/2/132}

\bibitem[{M. {Haywood} {et~al.}(2013){Haywood}, {Di Matteo}, {Lehnert}, {Katz},
  \& {G{\'o}mez}}]{Haywood13}
{Haywood}, M., {Di Matteo}, P., {Lehnert}, M.~D., {Katz}, D., \& {G{\'o}mez},
  A. 2013, \bibinfo{title}{{The age structure of stellar populations in the
  solar vicinity. Clues of a two-phase formation history of the Milky Way
  disk},} \aap, 560, A109, \dodoi{10.1051/0004-6361/201321397}

\bibitem[{M. {Haywood} {et~al.}(2018){Haywood}, {Di Matteo}, {Lehnert},
  {Snaith}, {Khoperskov}, \& {G{\'o}mez}}]{Haywood18}
{Haywood}, M., {Di Matteo}, P., {Lehnert}, M.~D., {et~al.} 2018,
  \bibinfo{title}{{In Disguise or Out of Reach: First Clues about In Situ and
  Accreted Stars in the Stellar Halo of the Milky Way from Gaia DR2},} \apj,
  863, 113, \dodoi{10.3847/1538-4357/aad235}

\bibitem[{A. {Helmi} {et~al.}(2018){Helmi}, {Babusiaux}, {Koppelman},
  {Massari}, {Veljanoski}, \& {Brown}}]{Helmi18}
{Helmi}, A., {Babusiaux}, C., {Koppelman}, H.~H., {et~al.} 2018,
  \bibinfo{title}{{The merger that led to the formation of the Milky Way's
  inner stellar halo and thick disk},} \nat, 563, 85,
  \dodoi{10.1038/s41586-018-0625-x}

\bibitem[{V. {Hill} {et~al.}(2019){Hill}, {Sk{\'u}lad{\'o}ttir}, {Tolstoy},
  {Venn}, {Shetrone}, {Jablonka}, {Primas}, {Battaglia}, {de Boer},
  {Fran{\c{c}}ois}, {Helmi}, {Kaufer}, {Letarte}, {Starkenburg}, \&
  {Spite}}]{Hill19}
{Hill}, V., {Sk{\'u}lad{\'o}ttir}, {\'A}., {Tolstoy}, E., {et~al.} 2019,
  \bibinfo{title}{{VLT/FLAMES high-resolution chemical abundances in Sculptor:
  a textbook dwarf spheroidal galaxy},} \aap, 626, A15,
  \dodoi{10.1051/0004-6361/201833950}

\bibitem[{D. {Horta} {et~al.}(2021){Horta}, {Schiavon}, {Mackereth}, {Pfeffer},
  {Mason}, {Kisku}, {Fragkoudi}, {Allende Prieto}, {Cunha}, {Hasselquist},
  {Holtzman}, {Majewski}, {Nataf}, {O'Connell}, {Schultheis}, \&
  {Smith}}]{Horta2021}
{Horta}, D., {Schiavon}, R.~P., {Mackereth}, J.~T., {et~al.} 2021,
  \bibinfo{title}{{Evidence from APOGEE for the presence of a major building
  block of the halo buried in the inner Galaxy},} \mnras, 500, 1385,
  \dodoi{10.1093/mnras/staa2987}

\bibitem[{R. {Ibata} {et~al.}(2021){Ibata}, {Malhan}, {Martin}, {Aubert},
  {Famaey}, {Bianchini}, {Monari}, {Siebert}, {Thomas}, {Bellazzini},
  {Bonifacio}, {Caffau}, \& {Renaud}}]{Ibata21}
{Ibata}, R., {Malhan}, K., {Martin}, N., {et~al.} 2021,
  \bibinfo{title}{{Charting the Galactic Acceleration Field. I. A Search for
  Stellar Streams with Gaia DR2 and EDR3 with Follow-up from ESPaDOnS and
  UVES},} \apj, 914, 123, \dodoi{10.3847/1538-4357/abfcc2}

\bibitem[{I. {Jean-Baptiste} {et~al.}(2017){Jean-Baptiste}, {Di Matteo},
  {Haywood}, {G{\'o}mez}, {Montuori}, {Combes}, \& {Semelin}}]{Jean-Baptiste17}
{Jean-Baptiste}, I., {Di Matteo}, P., {Haywood}, M., {et~al.} 2017,
  \bibinfo{title}{{On the kinematic detection of accreted streams in the Gaia
  era: a cautionary tale},} \aap, 604, A106,
  \dodoi{10.1051/0004-6361/201629691}

\bibitem[{K.~V. {Johnston} {et~al.}(1995){Johnston}, {Spergel}, \&
  {Hernquist}}]{Johnston95}
{Johnston}, K.~V., {Spergel}, D.~N., \& {Hernquist}, L. 1995,
  \bibinfo{title}{{The Disruption of the Sagittarius Dwarf Galaxy},} \apj, 451,
  598, \dodoi{10.1086/176247}

\bibitem[{A.~I. {Karakas} \& J.~C. {Lattanzio}(2014){Karakas} \&
  {Lattanzio}}]{Karakas14}
{Karakas}, A.~I., \& {Lattanzio}, J.~C. 2014, \bibinfo{title}{{The Dawes Review
  2: Nucleosynthesis and Stellar Yields of Low- and Intermediate-Mass Single
  Stars},} \pasa, 31, e030, \dodoi{10.1017/pasa.2014.21}

\bibitem[{S. {Khoperskov} {et~al.}(2023{\natexlab{a}}){Khoperskov}, {Minchev},
  {Libeskind}, {Haywood}, {Di Matteo}, {Belokurov}, {Steinmetz}, {Gomez},
  {Grand}, {Hoffman}, {Knebe}, {Sorce}, {Spaare}, {Tempel}, \&
  {Vogelsberger}}]{Khoperskov23b}
{Khoperskov}, S., {Minchev}, I., {Libeskind}, N., {et~al.} 2023{\natexlab{a}},
  \bibinfo{title}{{The stellar halo in Local Group Hestia simulations. II. The
  accreted component},} \aap, 677, A90, \dodoi{10.1051/0004-6361/202244233}

\bibitem[{S. {Khoperskov} {et~al.}(2023{\natexlab{b}}){Khoperskov}, {Minchev},
  {Libeskind}, {Haywood}, {Di Matteo}, {Belokurov}, {Steinmetz}, {Gomez},
  {Grand}, {Hoffman}, {Knebe}, {Sorce}, {Spaare}, {Tempel}, \&
  {Vogelsberger}}]{Khoperskov23a}
{Khoperskov}, S., {Minchev}, I., {Libeskind}, N., {et~al.} 2023{\natexlab{b}},
  \bibinfo{title}{{The stellar halo in Local Group Hestia simulations. II. The
  accreted component},} \aap, 677, A90, \dodoi{10.1051/0004-6361/202244233}

\bibitem[{E.~N. {Kirby} {et~al.}(2019){Kirby}, {Xie}, {Guo}, {de los Reyes},
  {Bergemann}, {Kovalev}, {Shen}, {Piro}, \& {McWilliam}}]{Kirby19}
{Kirby}, E.~N., {Xie}, J.~L., {Guo}, R., {et~al.} 2019,
  \bibinfo{title}{{Evidence for Sub-Chandrasekhar Type Ia Supernovae from
  Stellar Abundances in Dwarf Galaxies},} \apj, 881, 45,
  \dodoi{10.3847/1538-4357/ab2c02}

\bibitem[{C. {Kobayashi} {et~al.}(2020){Kobayashi}, {Karakas}, \&
  {Lugaro}}]{Kobayashi20}
{Kobayashi}, C., {Karakas}, A.~I., \& {Lugaro}, M. 2020, \bibinfo{title}{{The
  Origin of Elements from Carbon to Uranium},} \apj, 900, 179,
  \dodoi{10.3847/1538-4357/abae65}

\bibitem[{H.~H. {Koppelman} {et~al.}(2019){Koppelman}, {Helmi}, {Massari},
  {Price-Whelan}, \& {Starkenburg}}]{Koppelman19}
{Koppelman}, H.~H., {Helmi}, A., {Massari}, D., {Price-Whelan}, A.~M., \&
  {Starkenburg}, T.~K. 2019, \bibinfo{title}{{Multiple retrograde substructures
  in the Galactic halo: A shattered view of Galactic history},} \aap, 631, L9,
  \dodoi{10.1051/0004-6361/201936738}

\bibitem[{I. {Koutsouridou} {et~al.}(2025){Koutsouridou},
  {Sk{\'u}lad{\'o}ttir}, \& {Salvadori}}]{Koutsouridou25}
{Koutsouridou}, I., {Sk{\'u}lad{\'o}ttir}, {\'A}., \& {Salvadori}, S. 2025,
  \bibinfo{title}{{Large databases of metal-poor stars corrected for
  three-dimensional and/or non-local thermodynamic equilibrium effects},} arXiv
  e-prints, arXiv:2505.13607.
\newblock \doarXiv{2505.13607}

\bibitem[{S.~R. {Majewski} {et~al.}(2017){Majewski}, {Schiavon}, {Frinchaboy},
  {Allende Prieto}, {Barkhouser}, {Bizyaev}, {Blank}, {Brunner}, {Burton},
  {Carrera}, {Chojnowski}, {Cunha}, {Epstein}, {Fitzgerald}, {Garc{\'\i}a
  P{\'e}rez}, {Hearty}, {Henderson}, {Holtzman}, {Johnson}, {Lam}, {Lawler},
  {Maseman}, {M{\'e}sz{\'a}ros}, {Nelson}, {Nguyen}, {Nidever}, {Pinsonneault},
  {Shetrone}, {Smee}, {Smith}, {Stolberg}, {Skrutskie}, {Walker}, {Wilson},
  {Zasowski}, {Anders}, {Basu}, {Beland}, {Blanton}, {Bovy}, {Brownstein},
  {Carlberg}, {Chaplin}, {Chiappini}, {Eisenstein}, {Elsworth}, {Feuillet},
  {Fleming}, {Galbraith-Frew}, {Garc{\'\i}a}, {Garc{\'\i}a-Hern{\'a}ndez},
  {Gillespie}, {Girardi}, {Gunn}, {Hasselquist}, {Hayden}, {Hekker}, {Ivans},
  {Kinemuchi}, {Klaene}, {Mahadevan}, {Mathur}, {Mosser}, {Muna}, {Munn},
  {Nichol}, {O'Connell}, {Parejko}, {Robin}, {Rocha-Pinto}, {Schultheis},
  {Serenelli}, {Shane}, {Silva Aguirre}, {Sobeck}, {Thompson}, {Troup},
  {Weinberg}, \& {Zamora}}]{Majewski17}
{Majewski}, S.~R., {Schiavon}, R.~P., {Frinchaboy}, P.~M., {et~al.} 2017,
  \bibinfo{title}{{The Apache Point Observatory Galactic Evolution Experiment
  (APOGEE)},} \aj, 154, 94, \dodoi{10.3847/1538-3881/aa784d}

\bibitem[{T. {Matsuno} {et~al.}(2024){Matsuno}, {Amarsi}, {Carlos}, \&
  {Nissen}}]{Matsuno24}
{Matsuno}, T., {Amarsi}, A.~M., {Carlos}, M., \& {Nissen}, P.~E. 2024,
  \bibinfo{title}{{3D non-local thermodynamic equilibrium magnesium abundances
  reveal a distinct halo population},} \aap, 688, A72,
  \dodoi{10.1051/0004-6361/202450057}

\bibitem[{A. {Mori} {et~al.}(2024){Mori}, {Di Matteo}, {Salvadori},
  {Khoperskov}, {Pagnini}, \& {Haywood}}]{Mori24}
{Mori}, A., {Di Matteo}, P., {Salvadori}, S., {et~al.} 2024,
  \bibinfo{title}{{Metallicity distributions of halo stars: do they trace the
  Galactic accretion history?},} \aap, 690, A136,
  \dodoi{10.1051/0004-6361/202449291}

\bibitem[{G.~C. {Myeong} {et~al.}(2022){Myeong}, {Belokurov}, {Aguado},
  {Evans}, {Caldwell}, \& {Bradley}}]{Myeong22}
{Myeong}, G.~C., {Belokurov}, V., {Aguado}, D.~S., {et~al.} 2022,
  \bibinfo{title}{{Milky Way's Eccentric Constituents with Gaia, APOGEE, and
  GALAH},} \apj, 938, 21, \dodoi{10.3847/1538-4357/ac8d68}

\bibitem[{P.~E. {Nissen} {et~al.}(2024){Nissen}, {Amarsi},
  {Sk{\'u}lad{\'o}ttir}, \& {Schuster}}]{Nissen24}
{Nissen}, P.~E., {Amarsi}, A.~M., {Sk{\'u}lad{\'o}ttir}, {\'A}., \& {Schuster},
  W.~J. 2024, \bibinfo{title}{{Abundances of iron-peak elements in accreted and
  in situ born Galactic halo stars},} \aap, 682, A116,
  \dodoi{10.1051/0004-6361/202348392}

\bibitem[{P.~E. {Nissen} \& W.~J. {Schuster}(1997){Nissen} \&
  {Schuster}}]{Nissen97}
{Nissen}, P.~E., \& {Schuster}, W.~J. 1997, \bibinfo{title}{{Chemical
  composition of halo and disk stars with overlapping metallicities.},} \aap,
  326, 751

\bibitem[{P.~E. {Nissen} \& W.~J. {Schuster}(2010){Nissen} \&
  {Schuster}}]{Nissen10}
{Nissen}, P.~E., \& {Schuster}, W.~J. 2010, \bibinfo{title}{{Two distinct halo
  populations in the solar neighborhood. Evidence from stellar abundance ratios
  and kinematics},} \aap, 511, L10, \dodoi{10.1051/0004-6361/200913877}

\bibitem[{P.~E. {Nissen} \& W.~J. {Schuster}(2011){Nissen} \&
  {Schuster}}]{Nissen11}
{Nissen}, P.~E., \& {Schuster}, W.~J. 2011, \bibinfo{title}{{Two distinct halo
  populations in the solar neighborhood. II. Evidence from stellar abundances
  of Mn, Cu, Zn, Y, and Ba},} \aap, 530, A15,
  \dodoi{10.1051/0004-6361/201116619}

\bibitem[{T. {Nordlander} \& K. {Lind}(2017){Nordlander} \&
  {Lind}}]{Nordlander17}
{Nordlander}, T., \& {Lind}, K. 2017, \bibinfo{title}{{Non-LTE aluminium
  abundances in late-type stars},} \aap, 607, A75,
  \dodoi{10.1051/0004-6361/201730427}

\bibitem[{G. {Pagnini} {et~al.}(2023){Pagnini}, {Di Matteo}, {Khoperskov},
  {Mastrobuono-Battisti}, {Haywood}, {Renaud}, \& {Combes}}]{Pagnini23}
{Pagnini}, G., {Di Matteo}, P., {Khoperskov}, S., {et~al.} 2023,
  \bibinfo{title}{{The distribution of globular clusters in kinematic spaces
  does not trace the accretion history of the host galaxy},} \aap, 673, A86,
  \dodoi{10.1051/0004-6361/202245128}

\bibitem[{B. {Plez}(2012){Plez}}]{Plez12}
{Plez}, B. 2012, \bibinfo{title}{{Turbospectrum: Code for spectral
  synthesis},}, Astrophysics Source Code Library, record ascl:1205.004

\bibitem[{{\'A}. {Sk{\'u}lad{\'o}ttir} {et~al.}(2020){Sk{\'u}lad{\'o}ttir},
  {Hansen}, {Choplin}, {Salvadori}, {Hampel}, \& {Campbell}}]{Skuladottir20a}
{Sk{\'u}lad{\'o}ttir}, {\'A}., {Hansen}, C.~J., {Choplin}, A., {et~al.} 2020,
  \bibinfo{title}{{Neutron-capture elements in dwarf galaxies. II. Challenges
  for the s- and i-processes at low metallicity},} \aap, 634, A84,
  \dodoi{10.1051/0004-6361/201937075}

\bibitem[{{\'A}. {Sk{\'u}lad{\'o}ttir} \& S.
  {Salvadori}(2020){Sk{\'u}lad{\'o}ttir} \& {Salvadori}}]{Skuladottir20}
{Sk{\'u}lad{\'o}ttir}, {\'A}., \& {Salvadori}, S. 2020,
  \bibinfo{title}{{Evidence for {\ensuremath{\gtrsim}}4 Gyr timescales of
  neutron star mergers from Galactic archaeology},} \aap, 634, L2,
  \dodoi{10.1051/0004-6361/201937293}

\bibitem[{N. {Storm} \& M. {Bergemann}(2023){Storm} \& {Bergemann}}]{Storm23}
{Storm}, N., \& {Bergemann}, M. 2023, \bibinfo{title}{{Observational
  constraints on the origin of the elements - VII. NLTE analysis of Y II lines
  in spectra of cool stars and implications for Y as a Galactic chemical
  clock},} \mnras, 525, 3718, \dodoi{10.1093/mnras/stad2488}

\bibitem[{E. {Tolstoy} {et~al.}(2009){Tolstoy}, {Hill}, \& {Tosi}}]{Tolstoy09}
{Tolstoy}, E., {Hill}, V., \& {Tosi}, M. 2009, \bibinfo{title}{{Star-Formation
  Histories, Abundances, and Kinematics of Dwarf Galaxies in the Local Group},}
  \araa, 47, 371, \dodoi{10.1146/annurev-astro-082708-101650}

\bibitem[{E. {Tolstoy} {et~al.}(2023){Tolstoy}, {Sk{\'u}lad{\'o}ttir},
  {Battaglia}, {Brown}, {Massari}, {Irwin}, {Starkenburg}, {Salvadori}, {Hill},
  {Jablonka}, {Salaris}, {van Essen}, {Olsthoorn}, {Helmi}, \&
  {Pritchard}}]{Tolstoy23}
{Tolstoy}, E., {Sk{\'u}lad{\'o}ttir}, {\'A}., {Battaglia}, G., {et~al.} 2023,
  \bibinfo{title}{{A 3D view of dwarf galaxies with Gaia and VLT/FLAMES. I. The
  Sculptor dwarf spheroidal},} \aap, 675, A49,
  \dodoi{10.1051/0004-6361/202245717}

\end{thebibliography}
\bibliographystyle{aasjournal}

\clearpage
\pagebreak

\appendix

\begin{figure}
\begin{center}
\includegraphics[width=0.4\hsize]{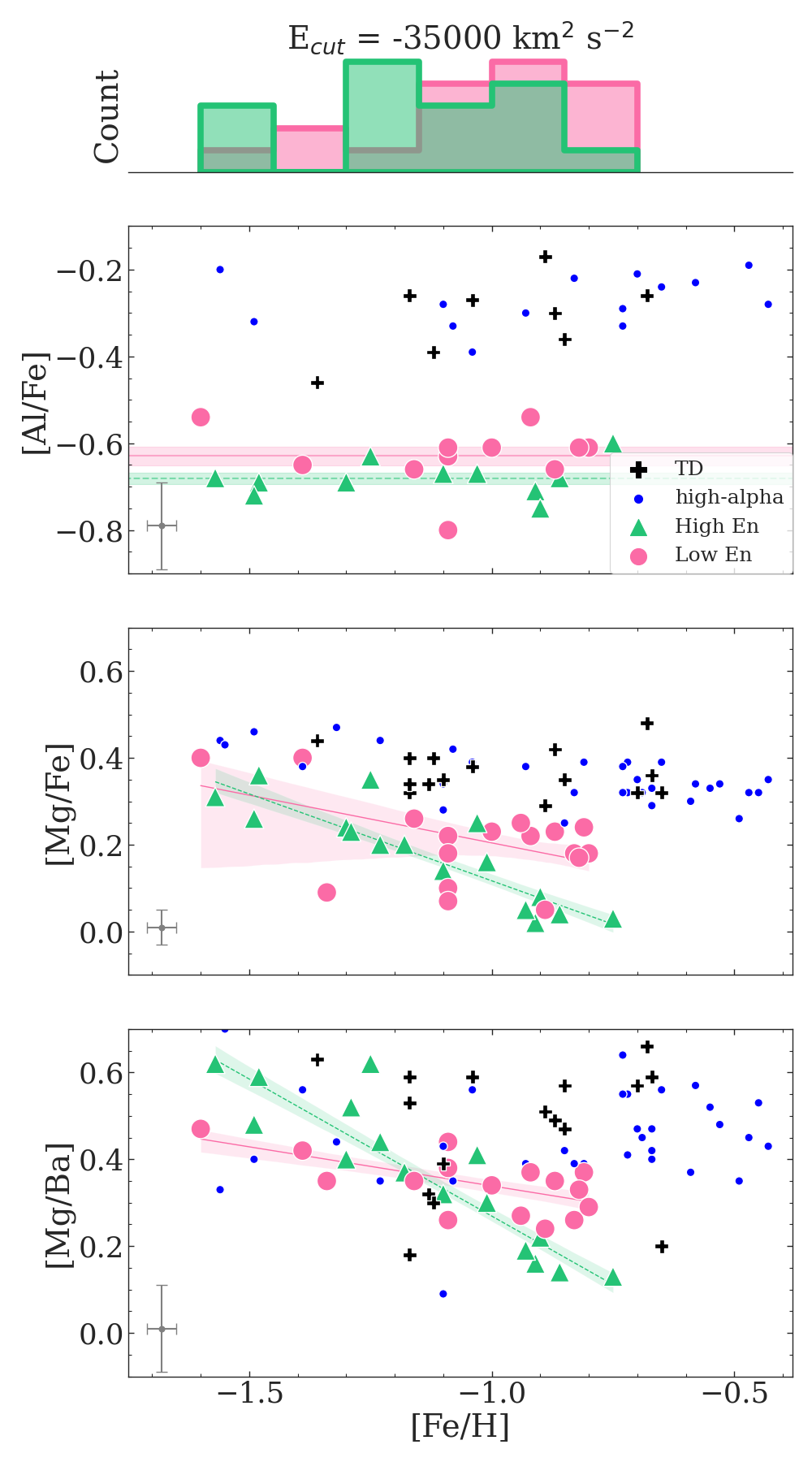} 
\includegraphics[width=0.4\hsize]{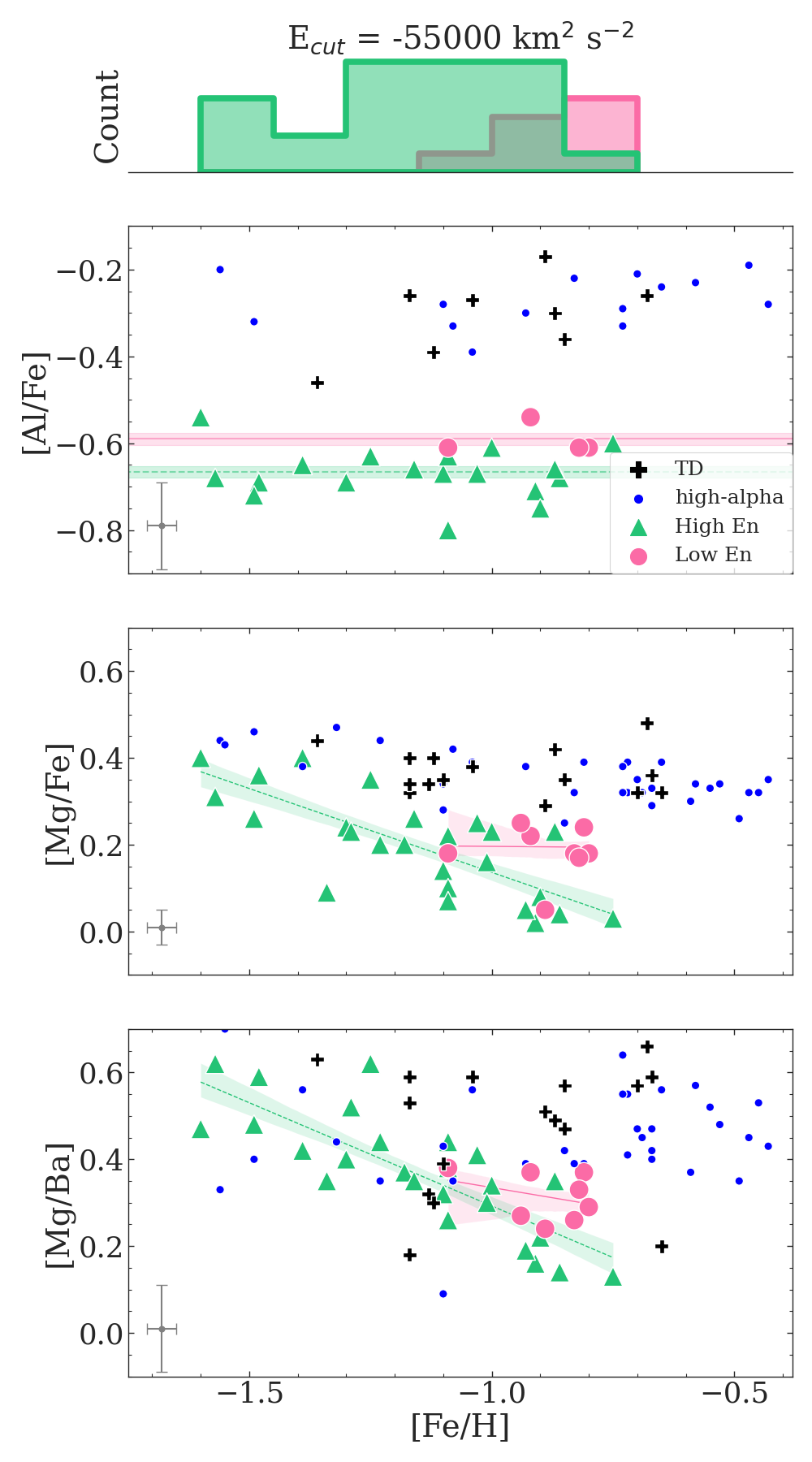} 
\caption{Same as Fig.~\ref{fig:abu} for different cuts in Energy: $E_{cut}=\rm -3.5\times10^4\,km^2\,s^{-2}$ (left), and $E_{cut}=\rm-5.5\times10^4\,km^2\,s^{-2}$ (right).
Chemical abundances for the accreted stellar sample with low (pink circles, $E<E_{cut}$) and high (green triangles, $E>E_{cut}$) energies. Lines in the [Al/Fe] panels show average values of the two subsamples, while shaded areas are the error of the mean. In the [Mg/Fe] and [Mg/Ba] panels, 
trend lines are shown with 68\% confidence interval. Top marginal plot shows the metallicity distribution of the accreted samples. For reference, the thick disk and high-$\alpha$ Milky Way populations created in situ are shown with small black crosses and blue circles, respectively. Representative error bars for individual stars are shown in the bottom left corner.
}
\label{fig:xabu}
\end{center}
\end{figure}

\begin{figure}
\begin{center}
\includegraphics[width=1\hsize]{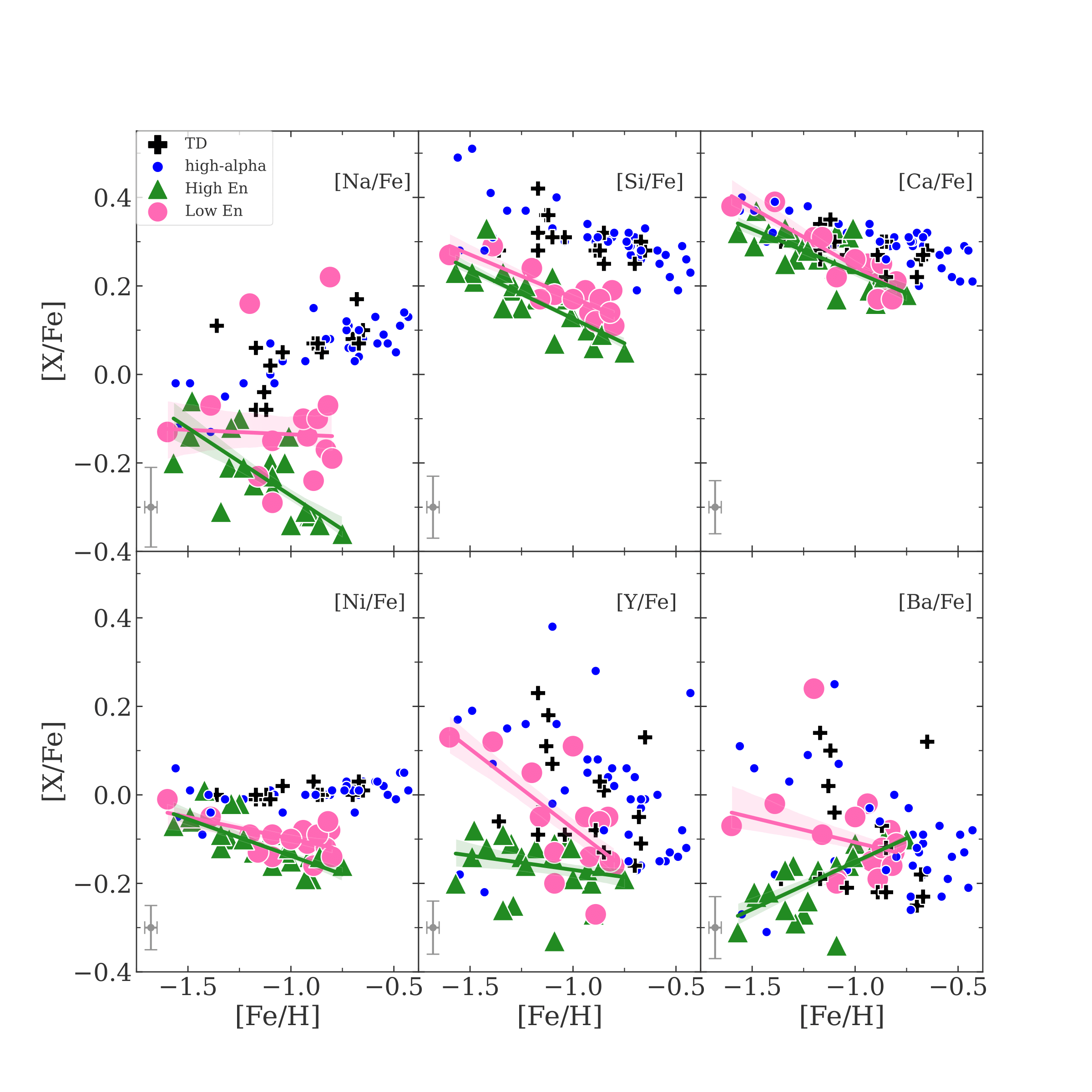} 
\caption{Chemical abundances for the NS accreted stellar sample with low (pink circles) and high (green triangles) energies, as defined in Fig.~\ref{fig:sel}.
For reference, the thick disk and high-$\alpha$ Milky Way populations created in situ are shown with small black crosses and blue circles, respectively. Representative error bars are shown in the bottom left corner of each panel.}
\label{fig:xabu}
\end{center}
\end{figure} 

\section{The Energy threshold $E_{cut}$}\label{sec:ecut}

Our fiducial energy threshold, $E_{cut}=-4.5\times10^4$\,km$^{2}$\,s$^{-1}$, is inspired by simulations, which show that stars above this limit should come from the outer regions of Gaia Enceladus (Fig.~\ref{fig:am}). It is important to understand how this choice of $E_{cut}$ affects our final conclusions (Sec.~\ref{sec:abu} and \ref{sec:con}). However, the range of reasonable $E_{cut}$ is limited by our sample size, which contains only 33 accreted stars which have both high-quality chemical abundances and energy measurements. Extensive tests have been done to assess the impact of changing the $E_{cut}$ and we found our results to be robust to this cut, within what limits are reasonable given our small sample size.

Fig.~\ref{fig:xabu} shows the abundance patterns from Fig.~\ref{fig:abu}, but with different $E_{cut}=-3.5\times10^4$\,km$^{2}$\,s$^{-1}$ (left), and $E_{cut}=-5.5\times10^4$\,km$^{2}$\,s$^{-1}$ (right), that is $\Delta E_{cut}=\pm10^4$\,km$^{2}$\,s$^{-1}$ compared to our fiducial value. In the case of [Al/Fe] the same qualitative trend is seen as in the fiducial case, where the low-energy sample has higher [Al/Fe] indicative of more efficient star formation (KS tests give $p<0.05$ for both $E_{cut}$). In the other abundance ratios, [Mg/Fe] and [Mg/Ba], the results are again qualitatively the same as from Fig.~\ref{fig:abu}, that is that the low-energy population (pink) consistently shows evidence of coming from an environment that experienced more efficient star formation, as expected in stars coming from the more inner parts of Gaia Enceladus, relative to the high-energy population (green). In all cases of [Mg/Fe] and [Mg/Ba], a KS test gives $p<0.01$ at $\rm[Fe/H]>-1$.

The MDFs of the high- and low-energy subsamples are, however, quite sensitive to $E_{cut}$. Moving $E_{cut}$ to lower energies (Fig.~\ref{fig:xabu}, right) will limit the low-energy sample only to quite high $\rm[Fe/H]\gtrsim-0.9$. In this case the two MDFs follow clearly distinct distributions ($p<0.01$). This is consistent with the simulations which predict that stars with lower kinetic energy come from the inner regions which are expected to be more metal-rich. In a similar way, moving the $E_{cut}$ to higher energies (Fig.~\ref{fig:xabu}, left), will make the MDF of the low-energy sample more metal-poor compared to the fiducial case shown in Fig.~\ref{fig:abu}. However, even though the MDFs are more similar in this case for the high- and low-energy subpopulations ($p=0.28$), the differences in the abundance patterns are still unambiguous, especially for [Mg/Fe] and [Mg/Ba] at high $\rm[Fe/H]>-1$. Thus, even when applying different $E_{cut}$ our conclusions hold: the subsample of stars with low (high) energy shows evidence of coming from a region of more (less) efficient star formation, consistent with originating from the inner (outer) regions of Gaia Enceladus.

\section{Additional Chemical Abundance Trends} \label{sec:xabu}

Fig.~\ref{fig:xabu} shows additional abundance ratios for the NS sample \citep{Nissen10,Nissen11,Nissen24}. Based on similar arguments as presented in Sec.~\ref{sec:data} we conservatively estimate the typical errors on individual stars to be: $\rm\Delta[Na/Fe]=0.09$, $\rm\Delta[Si/Fe]=0.07$,
$\rm\Delta[Ca/Fe]=0.06$,
$\rm\Delta[Ni/Fe]=0.05$,
$\rm\Delta[Y/Fe]=0.06$, and $\rm\Delta[Ba/Fe]=0.07$.

The abundance ratios of [Na/Fe], [Si/Fe], and [Ni/Fe] separate clearly in the high- and low-energy samples of accreted stars at high $\rm[Fe/H]\gtrsim-1$. This is consistent with the high-energy sample (green triangles), having formed in a region of less efficient star formation where the effect of the Fe production of SNIa becomes more enhanced \citep[e.g.][]{Tolstoy09}.\footnote{Although SNIa typically create substantial amounts of Ni \citep[e.g.][]{Kobayashi20}, this has been observationally shown to be not the case in low-metallicity systems with short star formation histories of only a few Gyr \citep[e.g.][]{Hill19,Kirby19}.} In the case of [Ca/Fe], however, the two samples converge and become indistinguishable. This can be explained by the fact that Ca is also created in significant amounts by SNIa \citep[e.g.][]{Kobayashi20}, so that this abundance ratio is less sensitive to small changes in the chemical enrichment efficiency. Finally, the abundance ratios of [Y/Fe] and [Ba/Fe] show clear differences in the high- (green triangles) versus low-energy accreted stars (pink circles). In both cases, the low-energy stars are more similar to the in-situ high-$\alpha$ population coming from regions of higher star formation efficiency compared to the accreted stars, indicating again that the low-energy sample experienced more efficient chemical enrichment compared to the high-energy sample.

In the case of Na, Si, and Ni over Fe, comparing the two distributions at $\rm[Fe/H]>-1$, gives $p<0.05$, while for Ca a KS test gives $p=0.5$. When comparing the $s$-process elements Y and Ba to Fe, instead the abundance trends merge at high [Fe/H] since both AGB stars and SNIa are delayed processes. However comparing the one-dimentional [Y/Fe] and [Ba/Fe] distributions over the full [Fe/H] range with a KS test gives $p<0.05$, confirming that statistically these two samples arise from two distinct distributions.
Overall, Fig.~\ref{fig:xabu} shows a clear distinction in the abundance trends of the high- and low-energy samples, confirming that the former was formed in an environment with less efficient star formation history compared to the latter.

\section{Table} \label{sec:table}

The new measurements of Al are listed in Table~\ref{tab:Al}, along with the Energies (E), and Angular momentum ($L_{z}$) for the entire NS sample. 

\begin{table}[]
%\resizebox{\textwidth}{!}{%
\tiny
    \centering
    \caption{Aluminum measurements for the NS sample, including $E$ and $L_z$.}
    \begin{tabular}{lcccccccccc}

\hline
\hline
ID & RA & DEC & E & L$_{z}$ & T$_{\rm eff}$ & log $g$ & v$_{t}$ & [Fe/H]$_{\rm 1D}$ & [Al/Fe]$_{3961\rm \AA}$ \\
NS24  & J2000 [deg] & J2000 [deg] &[km$^{2}$s$^{-2}$] & [km s$^{-1}$ kpc] & [K]  & & [km s$^{-1}$] & & &  \\
\hline
BD$-$21 3420 & 11 55 28.5 & $-$22 23 13 & $-$53645.085 & 1048.992   & 5909 & 4.30 & 1.12 & $-$1.14  & --- \\
CD$-$33 3337 & 06 54 47.8 & $-$33 44 49 & $-$40479.291 & 1443.189   & 6112 & 3.86 & 1.56 & $-$1.37  & $-$0.46 \\
CD$-$43 6810 & 11 08 40.1 & $-$44 15 34 & $-$56352.393 & 466.625    & 6059 & 4.32 & 1.24 & $-$0.44  & $-$0.28 \\
CD$-$45 3283 & 07 34 18.6 & $-$45 16 43 & $-$16310.167 & 15.251     & 5685 & 4.61 & 0.95 & $-$0.93  & $-$0.71 \\
CD$-$51 4628 & 10 17 14.9 & $-$52 29 19 & $-$16999.608 & 652.035    & 6296 & 4.29 & 1.31 & $-$1.32  & $-$0.69 \\
CD$-$57 1633 & 07 06 29.0 & $-$57 27 29 & $-$21327.224 & $-$138.987 & 5981 & 4.29 & 1.08 & $-$0.91  & $-$0.75 \\
CD$-$61 282  & 01 36 05.8 & $-$61 05 03 & $-$23761.893 & $-$479.358 & 5869 & 4.34 & 1.19 & $-$1.25  & --- \\
G05$-$19     & 03 11 26.5 & +12 37 10   & $-$15048.406 & $-$392.739 & 5970 & 4.28 & 1.17 & $-$1.19  & --- \\
G05$-$36     & 03 26 59.8 & +23 46 36   & $-$56067.378 & $-$921.575 & 6139 & 4.22 & 1.29 & $-$1.25  & --- \\
G05$-$40     & 03 27 39.4 & +21 02 35   & $-$49262.605 & $-$203.134 & 5892 & 4.20 & 1.12 & $-$0.83  & --- \\
G112$-$43    & 07 43 44.0 & $-$00 04 01 & $-$15903.390 & 1019.177   & 6209 & 4.02 & 1.17 & $-$1.27  & $-$0.63 \\
G112$-$44    & 07 43 44.1 & $-$00 03 49 & $-$15346.314 & 1010.833   & 5936 & 4.28 & 1.10 & $-$1.31  & --- \\
G114$-$42    & 09 10 44.9 & $-$03 48 09 &  11665.795   & 138.608    & 5721 & 4.40 & 1.19 & $-$1.12  & $-$0.67 \\
G119$-$64    & 11 12 48.0 & +35 43 44   & $-$29787.878 & $-$482.640 & 6333 & 4.14 & 1.40 & $-$1.50  & $-$0.69 \\
G121$-$12    & 11 44 35.7 & +25 32 12   & $-$5867.819  & 235.254    & 6041 & 4.25 & 1.26 & $-$0.94  & --- \\
G127$-$26    & 22 23 49.1 & +24 23 33   & $-$43184.521 & 1167.660   & 5886 & 4.20 & 1.11 & $-$0.53  & --- \\
G15$-$23     & 15 29 31.7 & +06 08 50   & $-$64435.591 & 239.553    & 5373 & 4.63 & 0.90 & $-$1.12  & --- \\
G150$-$40    & 13 48 52.1 & +27 40 10   & $-$62031.181 & $-$647.568 & 6080 & 4.11 & 1.31 & $-$0.82  & --- \\
G159$-$50    & 02 14 40.3 & $-$01 12 05 & $-$44901.473 & 258.338    & 5713 & 4.44 & 1.03 & $-$0.94  & $-$0.30 \\
G161$-$73    & 09 45 37.8 & $-$04 40 28 & $-$38250.425 & $-$251.232 & 6108 & 3.99 & 1.26 & $-$1.01  & $-$0.61 \\
G170$-$56    & 17 38 15.6 & +18 33 25   & $-$63192.086 & $-$401.160 & 6112 & 4.11 & 1.39 & $-$0.94  & $-$0.54 \\
G176$-$53    & 11 46 35.2 & +50 52 55   & $-$41545.126 & $-$255.758 & 5615 & 4.52 & 0.90 & $-$1.36  & --- \\
G18$-$28     & 22 05 40.7 & +12 22 36   & $-$66196.652 & $-$112.938 & 5443 & 4.49 & 0.88 & $-$0.85  & $-$0.22 \\
G18$-$39     & 22 18 36.5 & +08 26 45   & $-$50286.255 & $-$259.646 & 6112 & 4.23 & 1.44 & $-$1.41  & $-$0.65 \\
G180$-$24    & 16 03 13.3 & +42 14 47   & $-$59870.074 & $-$183.973 & 6137 & 4.20 & 1.45 & $-$1.41  & --- \\
G187$-$18    & 21 03 06.1 & +29 28 56   & $-$52583.489 & 702.202    & 5691 & 4.46 & 1.05 & $-$0.68  & --- \\
G188$-$22    & 21 43 57.1 & +27 23 24   & $-$42359.532 & 999.888    & 6116 & 4.20 & 1.42 & $-$1.33  & --- \\
G20$-$15     & 17 47 28.0 & $-$08 46 48 & $-$22673.616 & 1330.379   & 6162 & 4.32 & 1.50 & $-$1.50  & $-$0.72 \\
G21$-$22     & 18 39 09.7 & +00 07 14   & $-$38092.349 & $-$134.065 & 6021 & 4.27 & 1.30 & $-$1.10  & --- \\
G24$-$13     & 20 20 24.6 & +06 01 53   & $-$28451.921 & 1441.156   & 5764 & 4.38 & 0.86 & $-$0.73  & --- \\
G31$-$55     & 00 29 26.7 & $-$02 20 57 & $-$64420.910 & 326.309    & 5731 & 4.35 & 1.26 & $-$1.12  & $-$0.28 \\
G46$-$31     & 09 17 04.0 & +03 01 30   & $-$58525.963 & $-$818.054 & 6017 & 4.29 & 1.30 & $-$0.83  & --- \\
G49$-$19     & 09 38 50.6 & +28 24 09   & $-$62555.195 & $-$286.159 & 5863 & 4.32 & 1.12 & $-$0.55  & --- \\
G56$-$30     & 11 21 35.1 & +18 11 45   & $-$61442.672 & $-$578.373 & 5935 & 4.29 & 1.22 & $-$0.90  & --- \\
G56$-$36     & 11 23 16.2 & +19 53 38   & $-$60700.657 & $-$216.345 & 6067 & 4.33 & 1.33 & $-$0.94  & --- \\
G57$-$07     & 11 32 34.1 & +10 54 11   & $-$60998.951 & 367.302    & 5755 & 4.33 & 0.99 & $-$0.48  & $-$0.19 \\
G63$-$26     & 13 24 30.6 & +20 27 22   & $-$52353.783 &$-$1064.249 & 6175 & 4.17 & 1.65 & $-$1.58  & $-$0.20 \\
G66$-$22     & 14 43 18.0 & +05 49 40   & $-$25416.693 & 50.143     & 5297 & 4.46 & 0.78 & $-$0.88  & $-$0.68 \\
G74$-$32     & 02 34 13.2 & +33 00 05   & $-$51008.239 & 518.913    & 5864 & 4.41 & 1.04 & $-$0.74  & --- \\
G75$-$31     & 02 38 21.5 & +02 26 44   & $-$35.525    & 174.663    & 6135 & 4.02 & 1.28 & $-$1.04  & $-$0.67 \\
G81$-$02     & 04 03 55.3 & +39 44 19   & $-$48226.137 & $-$101.715 & 5967 & 4.24 & 1.21 & $-$0.69  & --- \\
G82$-$05     & 04 14 58.1 & $-$05 37 49 & $-$4999.966  & 382.635    & 5338 & 4.51 & 0.80 & $-$0.78  & $-$0.60 \\
G85$-$13     & 04 44 42.1 & +25 56 10   & $-$33526.502 & 1413.594   & 5709 & 4.46 & 0.87 & $-$0.60  & --- \\
G87$-$13     & 06 54 56.3 & +35 30 59   & $-$44208.041 & $-$418.805 & 6217 & 4.11 & 1.42 & $-$1.10  & $-$0.63 \\
G98$-$53     & 06 13 49.8 & +33 25 02   & $-$49878.986 & $-$205.329 & 5954 & 4.26 & 1.20 & $-$0.89  & $-$0.66 \\
G99$-$21     & 05 39 27.4 & +03 57 03   & $-$63425.139 & 251.187    & 5559 & 4.46 & 0.79 & $-$0.68  & --- \\
HD103723   & 11 56 36.0 & $-$21 25 10   & $-$62399.194 & 180.043    & 6050 & 4.20 & 1.11 & $-$0.81  & $-$0.61 \\
HD105004   & 12 05 24.9 & $-$26 35 44   & $-$62836.166 & $-$67.834  & 5852 & 4.35 & 1.09 & $-$0.83  & $-$0.61 \\
HD106516   & 12 15 10.6 & $-$10 18 45   & ---          & ---        & 6327 & 4.43 & 1.18 & $-$0.69  & $-$0.260 \\
HD111980   & 12 53 15.1 & $-$18 31 20   & $-$24046.511 & 193.329    & 5878 & 3.98 & 1.39 & $-$1.09  & $-$0.33 \\
HD113679   & 13 05 52.8 & $-$38 31 00   & $-$61093.063 & $-$456.275 & 5761 & 4.05 & 1.37 & $-$0.66  & $-$0.24 \\
HD114762   & 13 12 19.7 & +17 31 02     & $-$48233.467 & 1337.805   & 5956 & 4.24 & 1.37 & $-$0.72  & --- \\
HD120559   & 13 51 40.4 & $-$57 26 08   & $-$50937.033 & 1358.478   & 5486 & 4.58 & 1.05 & $-$0.91  & $-$0.17 \\
HD121004   & 13 53 58.1 & $-$46 32 20   & $-$58231.883 & $-$250.091 & 5755 & 4.43 & 1.16 & $-$0.71  & $-$0.21 \\
HD126681   & 14 27 24.9 & $-$18 24 40   & $-$47862.875 & 1434.020   & 5594 & 4.50 & 1.08 & $-$1.20  & --- \\
HD132475   & 14 59 49.8 & $-$22 00 46   & $-$54072.992 &$-$1101.602 & 5750 & 3.77 & 1.37 & $-$1.51  & $-$0.32 \\
HD148816   & 16 30 28.5 & +04 10 42     & $-$58060.148 & $-$328.621 & 5923 & 4.17 & 1.33 & $-$0.74  & $-$0.33 \\
HD159482   & 17 34 43.1 & +06 00 52     & $-$34441.619 & 1339.484   & 5829 & 4.37 & 1.21 & $-$0.74  & $-$0.29 \\
HD160693   & 17 39 36.9 & +37 11 02     & $-$30825.703 & 871.936    & 5809 & 4.35 & 1.02 & $-$0.48  & --- \\
HD163810   & 17 58 38.5 & $-$13 05 50   & ---          & ---        & 5592 & 4.61 & 1.17 & $-$1.22  & $-$0.69 \\
HD175179   & 18 54 23.2 & $-$04 36 19   & $-$55445.358 & 854.392    & 5804 & 4.40 & 1.08 & $-$0.66  & --- \\
HD17820    & 02 51 58.4 & +11 22 12     & $-$53944.090 & 1066.270   & 5873 & 4.28 & 1.27 & $-$0.68  & --- \\
HD179626   & 19 13 20.7 & $-$00 35 42   & $-$60088.291 & $-$80.795  & 5925 & 4.14 & 1.49 & $-$1.06  & $-$0.39 \\
HD189558   & 20 01 00.2 & $-$12 15 20   & $-$56007.779 & 895.495    & 5707 & 3.83 & 1.29 & $-$1.14  & $-$0.39 \\
HD193901   & 20 23 35.8 & $-$21 22 14   & $-$52271.971 & $-$254.695 & 5729 & 4.43 & 1.31 & $-$1.11  & $-$0.80 \\
HD194598   & 20 26 11.9 & +09 27 00     & $-$63379.593 & $-$334.626 & 6018 & 4.34 & 1.40 & $-$1.11  & $-$0.61 \\
HD199289   & 20 58 08.5 & $-$48 12 13   & $-$52282.125 & 1328.247   & 5915 & 4.30 & 1.21 & $-$1.05  & $-$0.27 \\
HD205650   & 21 37 26.0 & $-$27 38 07   & $-$50015.035 & 1098.811   & 5793 & 4.35 & 1.17 & $-$1.19  & $-$0.26 \\
HD222766   & 23 43 34.9 & $-$07 55 24   & $-$49723.834 & 327.322    & 5423 & 4.38 & 0.75 & $-$0.70  & --- \\
HD22879    & 03 40 22.1 & $-$03 13 01   & $-$50650.711 & 1133.472   & 5859 & 4.29 & 1.20 & $-$0.86  & $-$0.36 \\
HD230409   & 19 00 43.3 & +19 04 28     & $-$47369.250 & 853.548    & 5386 & 4.61 & 1.01 & $-$0.87  & --- \\
HD233511   & 08 19 22.6 & +54 05 10     & $-$56554.951 & $-$263.252 & 6125 & 4.21 & 1.20 & $-$1.58  & --- \\
HD237822   & 09 36 49.5 & +57 54 41     & $-$49912.862 & 724.422    & 5675 & 4.41 & 0.99 & $-$0.47  & --- \\
HD241253   & 05 09 57.0 & +05 33 27     & $-$51497.657 & 1106.419   & 5940 & 4.34 & 1.17 & $-$1.11  & --- \\
HD250792   & 06 03 14.9 & +19 21 39     & $-$29932.574 & $-$122.353 & 5572 & 4.50 & 0.98 & $-$1.03  & --- \\
HD25704    & 04 01 44.6 & $-$57 12 25   & $-$45495.201 & 1336.104   & 5974 & 4.30 & 1.33 & $-$0.86  & --- \\
HD284248   & 04 14 35.5 & +22 21 04     & $-$1604.685  & 509.746    & 6271 & 4.21 & 1.51 & $-$1.59  & $-$0.68 \\
HD3567     & 00 38 31.9 & $-$08 18 33   & $-$51702.450 & $-$243.333 & 6180 & 4.01 & 1.40 & $-$1.17  & $-$0.66 \\
HD51754    & 06 58 38.5 & $-$00 28 50   & $-$36020.813 & 565.665    & 5857 & 4.35 & 1.30 & $-$0.58  & $-$0.23 \\
HD59392    & 07 28 03.2 & $-$38 00 41   & $-$50976.824 & $-$798.023 & 6137 & 3.88 & 1.73 & $-$1.62  & $-$0.54 \\
HD76932    & 08 58 43.9 & $-$16 07 58   & $-$52726.972 & 1134.070   & 5977 & 4.17 & 1.30 & $-$0.87  & $-$0.30 \\
HD97320    & 11 11 00.7 & $-$65 25 38   & $-$40548.049 & 1662.606   & 6136 & 4.20 & 1.46 & $-$1.18  & --- \\
\hline
\hline
    \end{tabular}
    
    \label{tab:Al}
\end{table}

\clearpage
\pagebreak

\end{document}